\documentclass[fleqn,usenatbib]{mnras}

\usepackage{newtxtext,newtxmath}
\usepackage[T1]{fontenc}
\usepackage{ae,aecompl}

\usepackage{graphicx}	% Including figure files
\usepackage{amsmath}	% Advanced maths commands
\usepackage{amssymb}	% Extra maths symbols
\usepackage{booktabs,makecell}

\title[Statistical properties of Fermi GBM GRBs' spectra]{Statistical properties of Fermi GBM GRBs' spectra}

\author[I. I. R\'acz et al.]
{Istv\'an I. R\'acz$^{1,2}$\thanks{E-mail: racz@complex.elte.hu},
Lajos G. Bal\'azs$^{2,4}$,
Istvan Horvath$^{3}$,
L. Viktor T\'oth$^{2,4}$,
\newauthor Zsolt Bagoly$^{1}$
\\
$^{1}$Department of Physics of Complex Systems, E\"{o}tv\"{o}s Lor\'{a}nd University, P\'{a}zm\'{a}ny P\'{e}ter s\'{e}t\'{a}ny 1/A, H-1117 Budapest, Hungary\\
$^{2}$Konkoly Observatory, Research Centre for Astronomy and Earth Sciences, Hungarian Academy of Sciences,\\Konkoly Thege Mikl\'{o}s \'{u}t 15-17 Budapest, H-1121, Hungary\\
$^{3}$National University of Public Service, Ludovika t\'er 2, H-1083 Budapest, Hungary\\
$^{4}$Department of Astronomy, E\"{o}tv\"{o}s Lor\'{a}nd University, P\'{a}zm\'{a}ny P\'{e}ter s\'{e}t\'{a}ny 1/A, H-1117 Budapest, Hungary
}

\date{Accepted XXX. Received YYY; in original form ZZZ}

\pubyear{2017}

\begin{document}
\label{firstpage}
\pagerange{\pageref{firstpage}--\pageref{lastpage}}
\maketitle

\begin{abstract}
Statistical studies of gamma-ray burst (GRB) spectra may result in important information on the physics of GRBs. The Fermi GBM catalog contains GRB parameters (peak energy, spectral indices, intensity) estimated fitting the gamma-ray SED of the total emission (fluence, \textit{flnc}), and during the time of the peak flux (\textit{pflx}). Using contingency tables we studied the relationship of the models best fitting \textit{pflx} and \textit{flnc} time intervals.Our analysis revealed an ordering of the spectra into a power law - Comptonized - smoothly broken power law - Band series. This result was further supported by a correspondence analysis (CA) of the \textit{pflx} and \textit{flnc} spectra categorical variables. We performed a linear discriminant analysis (LDA) to find a relationship between categorical (spectral) and model independent physical data. LDA resulted in highly significant physical differences among the spectral types, that is more pronounced in the case of the \textit{pflx} spectra, than for the \textit{flnc} spectra. We interpreted this difference as caused by the temporal variation of the spectrum during the outburst. This spectral variability is confirmed by the differences in the low energy spectral index and peak energy, between the \textit{pflx} and \textit{flnc} spectra. We found that the synchrotron radiation is significant in GBM spectra. The mean low energy spectral index is close to the canonical value of $\alpha = -2/3$ during the peak flux. However, $\alpha$ is $\sim -0.9$ for the spectra of the fluences. We interpret this difference as showing that the effect of cooling is important only for the fluence spectra.
\end{abstract}


%%%%%%%%%%%%%%%%%%%%%%%%%%%%%%%%%%%%%%%%%%%%%%%%%%

%%%%%%%%%%%%%%%%% BODY OF PAPER %%%%%%%%%%%%%%%%%%

\section{Introduction}

Spectral energy distribution (SED) of gamma ray bursts (GRBs), the most energetic transients in the Universe \citep{Kumar}, represents basic information on the physical processes responsible for the observed characteristics of these objects \citep{Kumar6}. In the gamma energy range the SED can be well approximated by a combination of a small number of power law functions \citep{grb_func, Kumar5}, indicating that nonthermal (synchrotron, inverse Compton) processes play probably a fundamental role in the radiation \citep{meszaros_fireball,prompt_emission,peter_meszaros_2015}. In particular cases, however, there are indications also for the presence of a thermal component in the SED \citep{thermal_components,Ryde01,Ryde02}.

The Fermi Gamma-ray Space Telescope (FGST, Fermi satellite) (launched on 2008 June 11 into a Low Earth orbit of ~565 km) provided the scientific community with a wealth of high quality data on several astrophysical objects, including GRBs in the 8 keV to 40 MeV (GBM) \citep{fermi_gbm1,fermi_gbm2, fermi2} and 20 MeV to 300 GeV (LAT) ranges \citep{fermi_lat_technical,fermi_lat}.

In several cases the Band function \citep{band, Kumar7} fits well the observed SED within the limits of the statistical inference. There are GRBs, however, at which even some simpler functional forms (power law, cutted power law, smoothly broken power law) can give satisfactory fits \citep{comptonized1,smothly,band_smoothly,comptonized2}. The analytic forms of the spectra differ from each other in the number of the fitted parameters. 
 The attempt to fit more complex spectra to the observed gamma energy distributions were not really successful in most cases \citep{thermal1,thermal3,thermal2}. However, the Band function at the bright long GRBs' prompt radiation is often inadequate to fit the data \citep{peter_meszaros_2014B}. Nevertheless, it is worth mentioning these analytic models do not really have a physical background, although, they can be used to interpret some physical parameters (e.g. low- and high-energy spectral index, or peak energy).

A single power law has two free parameters: the $A$ amplitude and the $\alpha$ spectral index. A pivot energy ($E_{\mbox{piv}}$) can be defined to normalize the model to the energy range under consideration. It was fixed at 100 keV for GRBs detected by GBM.

Band's GRB function is usually considered as the standard for fitting GRB spectra. This function has four free parameters: the $A$ amplitude, $\alpha$ and $\beta$ the low- and high energy spectral indices, respectively, and the $\mbox{E}_{\mbox{peak}}$ energy.

The Comptonized model, a subset of the Band function in the limit when $\beta \rightarrow \infty$, is an exponentially cutoff power law. It has three free parameters: the $A$ amplitude, the $\alpha$ low-energy spectral index and $\mbox{E}_{\mbox{peak}}$. Similarly to the power-law model $E_{piv}$ is again fixed at 100 keV.

Finally, the broken power law spectra is characterized by one break with flexible curvature and is able to fit spectra with sharp or smooth transitions between the low- and high-energy power laws. This model was introduced by \citet{smothly}.

Although the functional form of the SED provides important information for understanding the physical processes responsible \citep{Kumar2,Kumar3} for the observed gamma radiation information on the model independent quantities as the duration, fluence, flux also bear important physical information. In the following we try to find relationships between the spectral data and model independent quantities \citep{peter_meszaros_2014}.

The majority of GRBs has a break in the spectral distribution recorded by the GBM \citep{fermi1,fermi3}. The presence of an observable break also depends on the number of the recorded photons. In the case of faint sources a simple power law has a satisfactory fit in GBMs spectral range, within the limits of the statistical inference. 

The paper is organized as follows: in Section~\ref{speda} we give an overview on the Fermi data we use in our analysis. The mathematical tools used to study the statistical properties of the data are given in Section~\ref{mulan}. We deal with interpreting the statistical results in Section~\ref{disc}. Finally, Section~\ref{sucon} briefly summarizes the main results and conclusions of our paper.

\section{GBM Spectral Data}
\label{speda}

The Fermi GBM catalogue contains two types of spectra: one referring to the peak flux (designated with \textit{pflx}) and the other one to the fluence (designated with \textit{flnc}). The peak flux spectra relates to the photons collected in the time bin around the intensity peak of the prompt emission (the time bin equals to 1,024 ms at the bursts of $T90 > 2 s$ duration and 64 ms if $T90 < 2 s$). Both in the \textit{pflx} and \textit{flnc} spectra four different  model spectra were fitted to every burst event and the best fitting model was indicated by the likelihood-based statistic. The four fitted model spectra are the power law, the Comptonized, the Band and the smoothly broken power law.

The FERMIGRBST\footnote{\url{https://heasarc.gsfc.nasa.gov/W3Browse/fermi/fermigbrst.html}} catalog \citep{fermi4, fermi5} contains several physical parameters derived from the rough data delivered by the Fermi GBM detectors. The database used in our paper contains 2069 records (number of GRBs, recorded until 1th May of 2017). One set of the data (e.g. duration, peak flux, fluence) is related to the burst in general and the other ones were obtained from fitting the models, along with the best fitted one. In this work we used these FERMIGRBST catalog data. 

In our analysis we took into account only the best fitted models in both sets (\textit{pflx} and \textit{flnc}) of model spectra so we ignored the remaining three spectra and their estimated parameters. As for the model independent parameters we took into account the $T50$ and $T90$ durations, the 64/256/1024 ms flux in the 10---1000 keV energy range, the 64/256/1024 ms (BATSE standard) flux in the 50---300 keV energy range, the 10---1000 keV energy band fluence, and 50---300 keV energy band BATSE fluence. As for the model fitted parameters we were dealing with the $\mbox{E}_{\mbox{peak}}$ and Low-Energy indices of Band, Smoothly broken power law and Comptonized. 
In the following Section we used several multivariate statistical methods to study the statistical relationships using these data as inputs.

\section{Multivariate Analysis of the Data}
\label{mulan}

As we mentioned in the previous Section the quantities listed in the GBM catalog have different data types. A significant fraction of the data belongs to the type having non-missing numeric values (e.g. duration, flux, fluence).

The spectral types, however, are assigned to the best fit models of the peak flux and the fluence form two categorical variables. The cross tabulation of these spectral type categories results in a table called contingency table \citep{contingency}, representing in each cell the frequency of the joint incidence of any pairs of the spectral types assigned to the peak flux and fluence.

At first we investigated the relationship between the categorical variables representing the best fit to the SEDs of GRBs. After getting the relationship between the peak flux and fluence spectral categories we study the connections of the numeric variables to these categories.

\subsection{Analysis of contingency tables}
\label{ctanal}

The cross tabulation of peak flux and fluence spectral types available in the GBM catalogue was resulted in a contingency table shown in Table~\ref{contab}. It contains only bursts with fitted spectral types were taken into account.

\begin{table}
  \centering
  \begin{tabular}{rrrrrr}
  \hline
 & fband & fsbpl & fcomp & fplaw & row.sum \\ 
  \hline
  pband &  50 &   9 &   4 &   0 &  63 \\ 
  psbpl &  12 &  12 &   5 &   0 &  29 \\ 
  pcomp & 100 &  64 & 624 &  25 & 813 \\ 
  pplaw &  12 &  16 & 500 & 419 & 947 \\ 
  col.sum & 174 & 101 & 1133 & 444 & 1852 \\ 
   \hline
\end{tabular}
\caption{Contingency table of the spectral types obtained according to the peak flux (rows) and fluences (columns). The last row and column mark the total number of objects in rows and columns, respectively. The table contains only those bursts having fitted spectral types. The first character of the row names refer to the spectra obtained from peak fluxes and the ones starting with f are obtained from fluences.}\label{contab}
\end{table}

Out of the 2069 GRBs in the FERMIGBST catalog only 1852 have non-missing values in all of the parameters studied. The right bottom corner of the Table contains the grand total, the $N=1852$ number of GRBs taken into account at  tabulating the data. Dividing counts in the individual cells by this grand total one gets a probability table from the original cross tabulation.

The numbers of the cells in Table~\ref{protab} represent the $p_{ij}$ joint probability of getting the spectral type in the $i^{th}$ row with that in the $j^{th}$ column. By definition the relationship between the joint and the conditional probability is given by

\begin{equation}\label{pden}
    p_{ij} = p(i|j) c_j = p(j|i) r_i \phantom{@@@} (i,j=1,2,3,4)
\end{equation}

\noindent where $c_j$ and $r_i$ give the column margin and row margin in Table \ref{protab}, respectively.

\begin{table}
  \centering
\begin{tabular}{rrrrrr}
  \hline
 & fband & fsbpl & fcomp & fplaw & row.sum \\ 
  \hline
  pband & 0.03 & 0.00 & 0.00 & 0.00 & 0.03 \\ 
  psbpl & 0.01 & 0.01 & 0.00 & 0.00 & 0.02 \\ 
  pcomp & 0.05 & 0.03 & 0.34 & 0.01 & 0.44 \\ 
  pplaw & 0.01 & 0.01 & 0.27 & 0.23 & 0.51 \\ \hline
  col.sum & 0.09 & 0.05 & 0.61 & 0.24 & 1.00 \\ 
   \hline
\end{tabular}
\caption{Probability Table of the spectral data. This table was obtained from Table~\ref{contab} by dividing the numbers in the cells by the number of the objects in the sample (grand total). The numbers in the cells express the joint probability of getting a spectral type from the peak flux (given in the rows) and that from the fluence (given in the columns).}\label{protab}
\end{table}

\subsubsection{Similarity between columns and between rows}
\label{simcr}

Dividing $p_{ij}$ by the $c_j$ column margin one gets the $p(i|j)$ conditional probability. It expresses the chance that the best fitting  spectral type of a burst was $i$ obtained from the peak flux, assuming that the best fitting spectral type is $j$ obtained from the fluence. The set of the $p(i|j)$ values are called {\it column profiles}. The column profiles obtained in this way is shown in Table \ref{cpro}.

\begin{table}
\centering
\begin{tabular}{rrrrrr}
  \hline
   & fband & fsbpl & fcomp & fplaw & av.pro \\ 
  \hline
  pband & 0.29 & 0.09 & 0.00 & 0.00 & 0.03 \\ 
  psbpl & 0.07 & 0.12 & 0.00 & 0.00 & 0.02 \\ 
  pcomp & 0.57 & 0.63 & 0.55 & 0.06 & 0.44 \\ 
  pplaw & 0.07 & 0.16 & 0.44 & 0.94 & 0.51 \\ \hline
  col.sum & 1.00 & 1.00 & 1.00 & 1.00 & 1.00 \\ 
   \hline
\end{tabular}
\caption{Column profiles. Each column  represents the probabilities of spectral types (listed in the row names) obtained from the  peak flux assuming a spectral type of  the fluence, according to its column. If there were no associations between the spectral types obtained from the peak flux and the fluence all the profiles would be similar to the average profile. Note the significant deviation of the columns of Band and power law spectra from the average profile, indicating a strong association.} \label{cpro}
\end{table}

Similarly, we get $p(j|i)$ conditional probability, the row profile, dividing the $p_{ij}$ joint probability by the $r_i$ row margin in Equation~(\ref{pden}). The row profile gives the probabilities of spectral types obtained from the fluence assuming a given one estimated from the peak flux. The row profiles obtained in this procedure are given in Table~\ref{rpro}.

\begin{table}
\centering
\begin{tabular}{rrrrrr}
  \hline
   & fband & fsbpl & fcomp & fplaw & row.sum \\ 
  \hline
  pband & 0.79 & 0.14 & 0.06 & 0.00 & 1.00 \\ 
  psbpl & 0.41 & 0.41 & 0.17 & 0.00 & 1.00 \\ 
  pcomp & 0.12 & 0.08 & 0.77 & 0.03 & 1.00 \\ 
  pplaw & 0.01 & 0.02 & 0.53 & 0.44 & 1.00 \\ \hline
  av.pro & 0.09 & 0.05 & 0.61 & 0.24 & 1.00 \\ 
   \hline
\end{tabular}
\caption{Row profiles. Each row gives the probability of getting a spectral type from the fluences assuming a given spectral type of the observed peak flux. If there were no associations between the spectral types obtained from the peak flux and the fluence, all the profiles in the Table would be similar to the average profile. Note the significant deviation of the Band profile (the pband line in the Table) from the average one.} \label{rpro}
\end{table}

\begin{figure}
\centering
  \includegraphics[width=3.9cm]{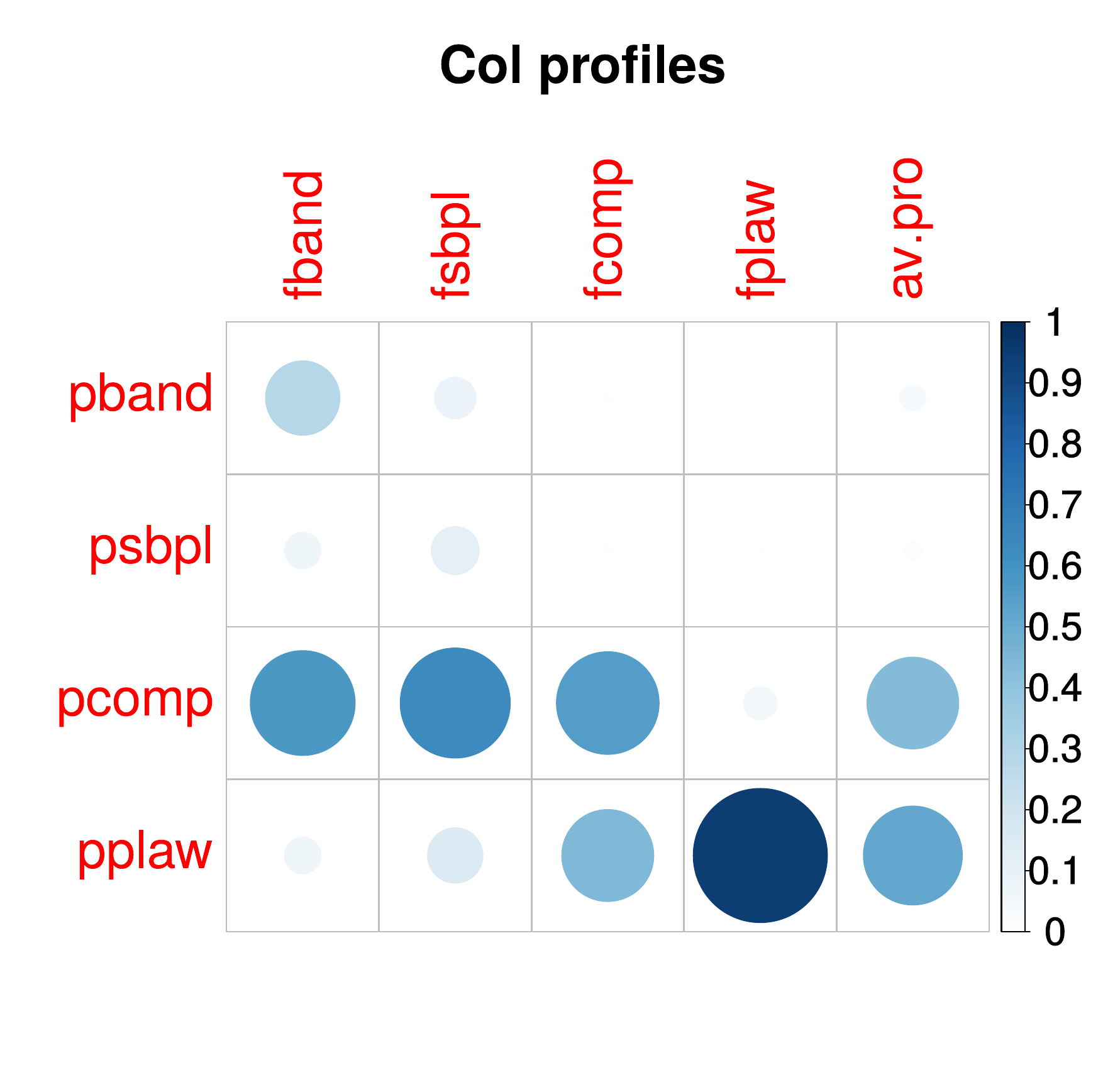} \hspace{2mm}
  \includegraphics[width=3.6cm]{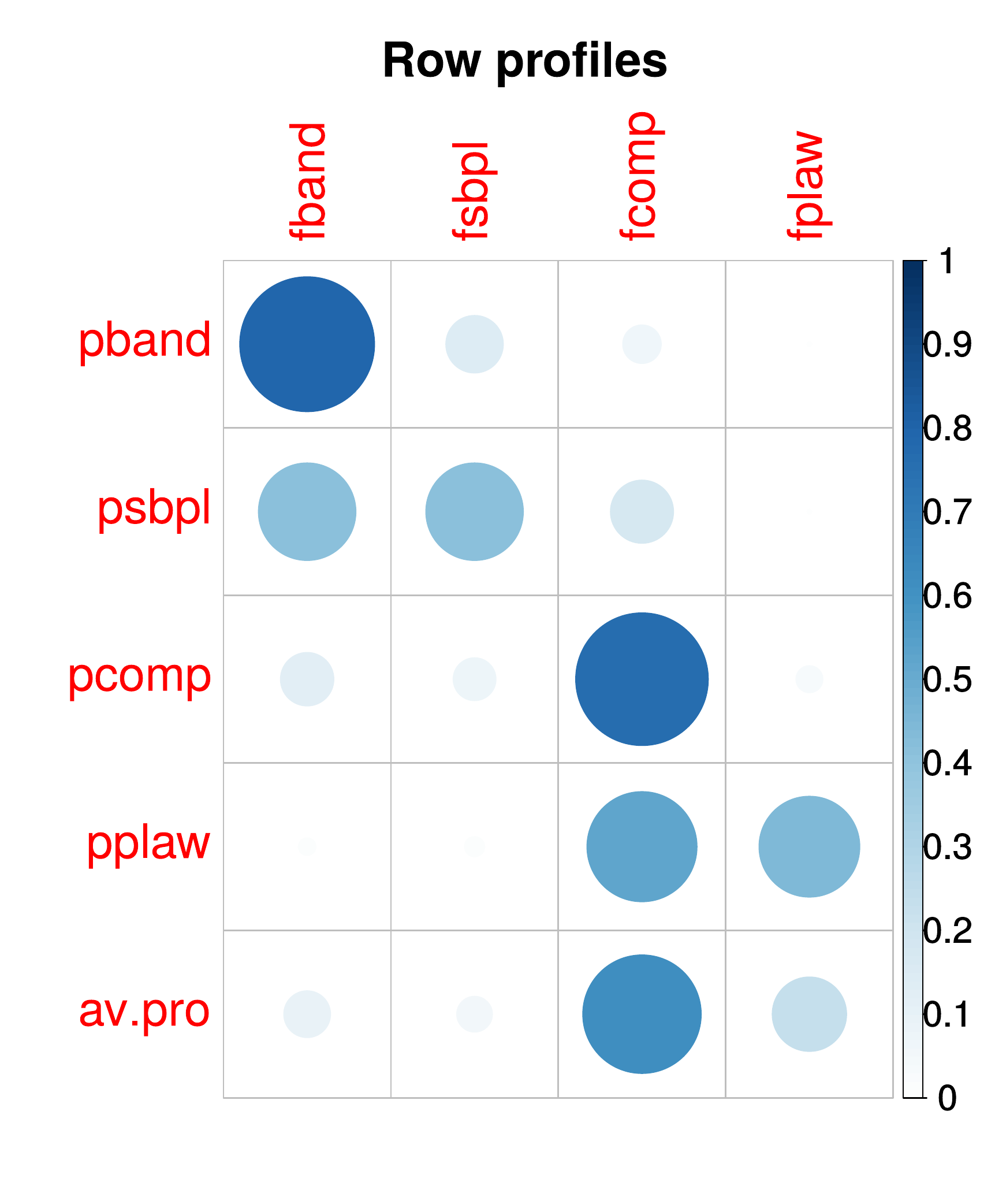}
\caption{Color display of the column (left panel) and row (right panel) profiles. The size and darkness of the circles represent the magnitude of the numeric value of the cells in Table~\ref{cpro}~and~\ref{rpro}. Note the significant difference between the first  and fourth  column profiles and the first and fourth row profiles (Band and power law in both cases).}\label{crdis}
\end{figure}

The properties of column and row profiles given in Tables \ref{cpro} and \ref{rpro} can be displayed in in Fig.~\ref{crdis}, in color form \citep{corrplot}. Inspecting this Figure one gets interesting rules. The spectra obtained from the peak flux precedes that from the fluence. The column profiles in Fig.~\ref{crdis} express the possible predecessors of a given spectrum of the fluence. In a similar way,  the row profiles indicate the probability of a succeeder fluence spectra of a given peak flux spectrum.

The column profiles in Fig.~\ref{crdis} indicate that a Comptonized spectrum of the peak flux can be a predecessor of a smoothly broken power law or Band type but a power law in the fluence only with a very low probability.

The row profiles in Fig.~\ref{crdis} show that a Band spectrum of the peak flux can evolve into the same type with very high probability. The probability, however, to get other spectral types is low, decreasing from smoothed power law through Comptonized spectrum to a power law predecessor at the end with a very low probability. On the other hand, a power law spectrum of the peak flux almost never proceeded into a Band spectrum in the fluence.

This means the spectral types can be ordered into a series of power law, Comptonized, smoothly broken power law and Band spectrum. A spectrum obtained by fitting a peak flux can evolve along this line by getting a spectrum of the fluence but with a very low probability backwards.

The similarity between column or row profiles can be expressed numerically by introducing squared Euclidean distances in the following way:

\begin{equation}\label{dispro}
    d^2(\mbox{profile}_1,\mbox{profile}_2)=\sum \frac{(\mbox{profile}_1-\mbox{profile}_2)^2}{\mbox{average profile}}.
\end{equation}

\begin{figure}
\centering
  \includegraphics[width=3.8cm]{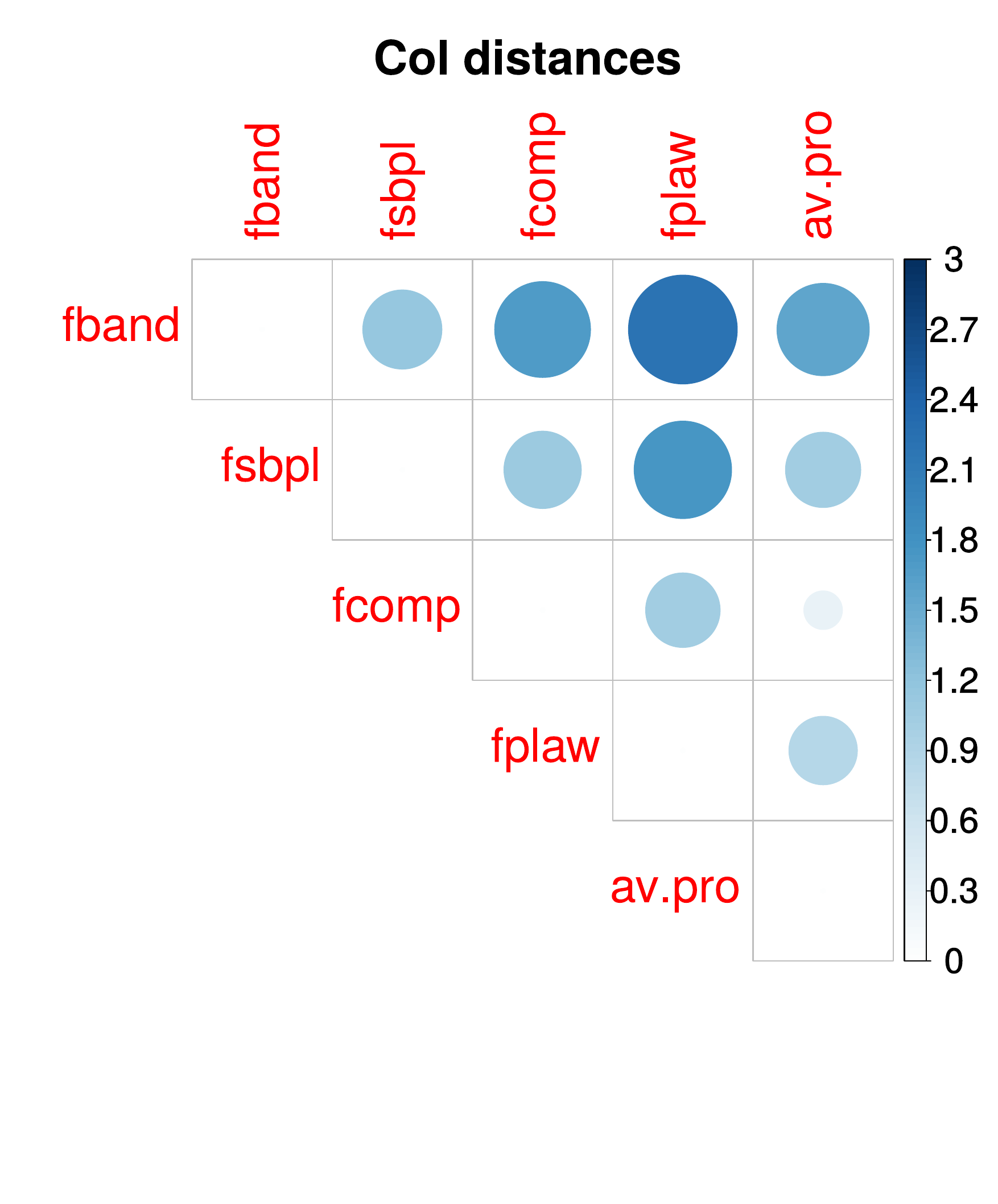} \hspace{2mm}
  \includegraphics[width=3.8cm]{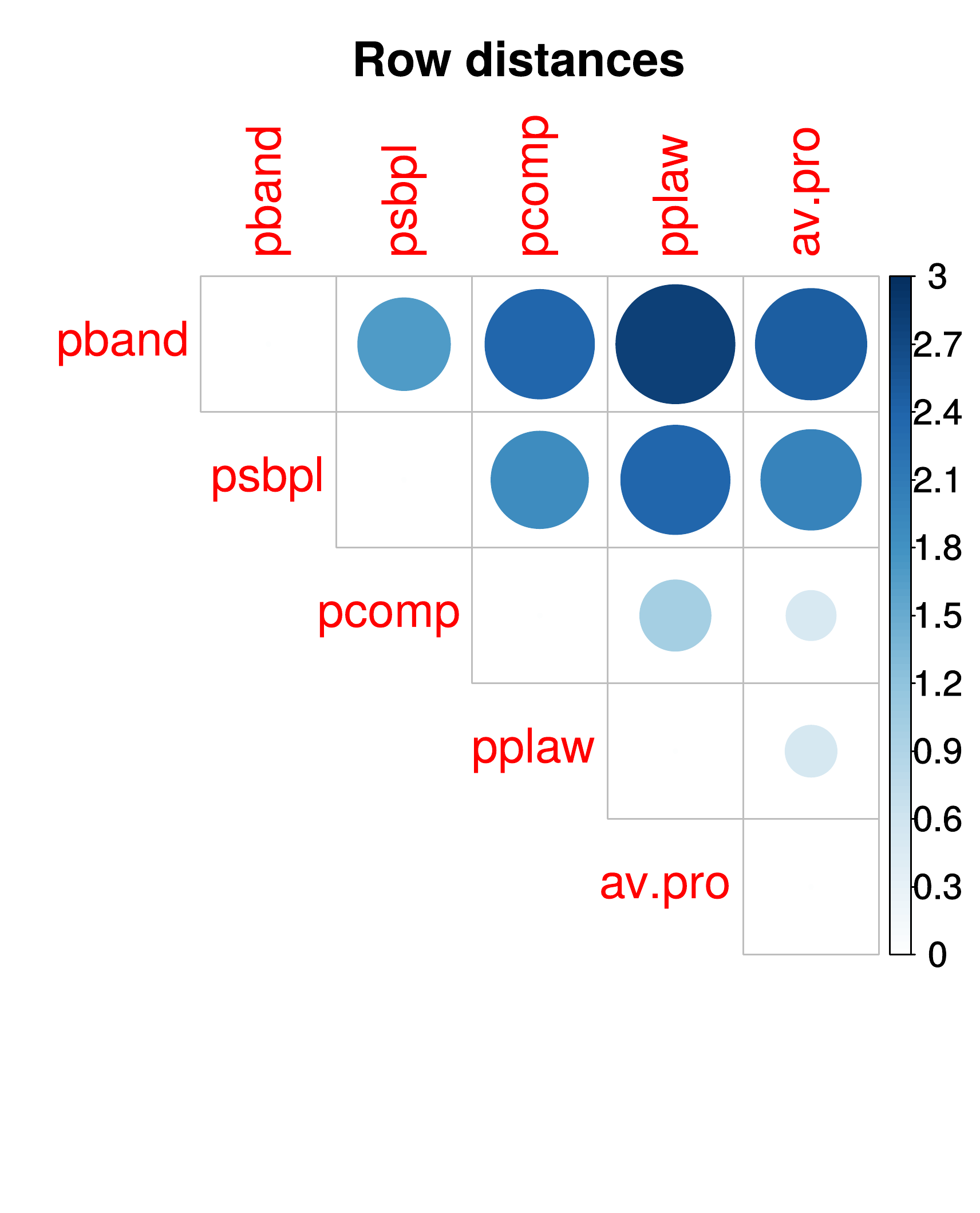}
 \caption{Euclidean distances between the column and row profiles. Note that the greatest distance is between the Band and power law types. The distances also indicate an order of the types: the power law, Comptonized spectra, smoothly broken power law and Band.}\label{crd2}
\end{figure}

\noindent The mutual distances of the column and row profiles are displayed in Fig.~\ref{crd2} in a colored form.

\subsubsection{Homogeneity test of the contingency table}
\label{homtab}

If there were no associations between the spectral types obtained from the peak fluxes and the fluences, the $p_{ij}$ joint probabilities in Equation~(\ref{pden}) would be equal to the $r_ic_j$ product of the row and column margins. The $n_{ij}$ cell values of Table~\ref{protab} would follow a Poisson equation with a $\lambda_{ij}=N r_ic_j$ parameter ($N$ is the sample size)

\begin{equation}\label{Õoi}
    P_{ij}(n) = \frac{\lambda_{ij}^n}{n!}exp(-\lambda_{ij})
\end{equation}

\noindent where $n$ is the numeric content of the actual cell. In the case of a Poisson distribution both the expected value and the variance equals to $ \lambda_{ij}$. Subtracting the expected value from the $n_{ij}$ cell content and dividing with the squared root of the $\lambda_{ij}$ we get the standardized (Pearson's) residuals of Table~\ref{contab}, the contingency table

\begin{equation}\label{stvar}
    s_{ij}=\frac{n_{ij}-\lambda_{ij}}{\sqrt{\lambda_{ij}}}
\end{equation}

\noindent where $s_{ij}$ has zero expected value and unit variance. Consequently, $s_{ij}^2$ follows a $\chi^2$ distribution and

\begin{equation}\label{chi2}
    \chi^2_{df}=\sum \limits^{nr}_{i=1}\sum \limits^{nc}_{j=1}
    s^2_{ij}
\end{equation}

\noindent is a $\chi^2_{df}$ distribution of $df=(nr-1)(nc-1)$  degrees of freedom ($nr$ and $nc$ are the numbers of rows and columns in Table~\ref{contab}).

The $\chi^2_{df}$ variable at the left side of Equation~(\ref{chi2}) enables us to test the homogeneity of the contingency table. As an $H_0$ null hypothesis we assume that there is no real association between the spectral types obtained from the peak flux and the fluence. If the $\chi^2_{df}$ value we obtained is only a random fluctuation, the contingency table is homogeneous.

Applying the {\tt chisq.test()} procedure for Table~\ref{contab} available in the R \citep{R} package, we obtained $\chi^2_{df}$~=~856.34 with a $p < 2.2\cdot10^{-16}$ probability for getting this value only accidentally. So we can reject the $H_0$ null hypothesis of homogeneity at a very high level of significance.

\begin{figure}
  \centering
  \includegraphics[width=3.8cm]{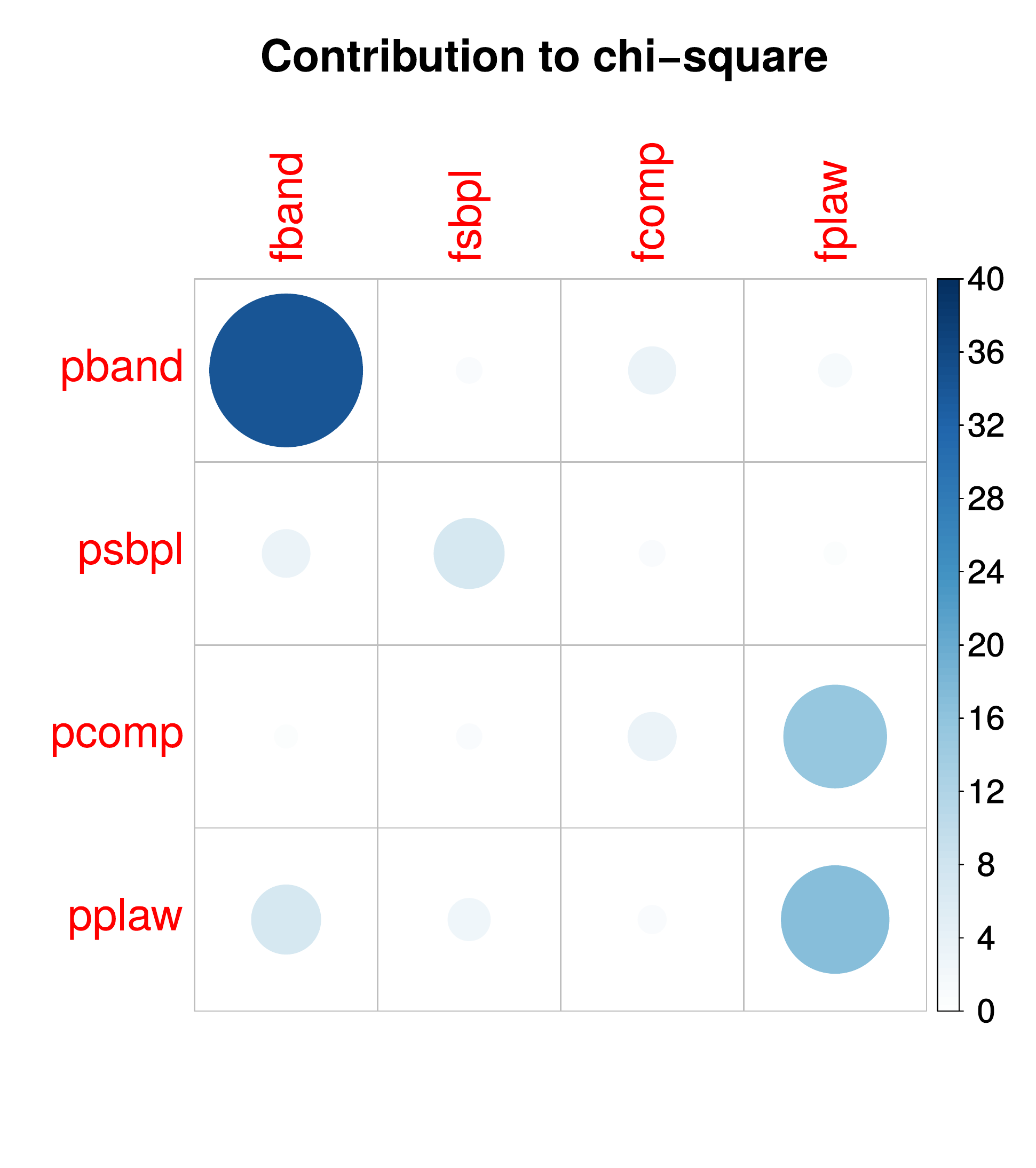} \hspace{2mm}
  \includegraphics[width=3.8cm]{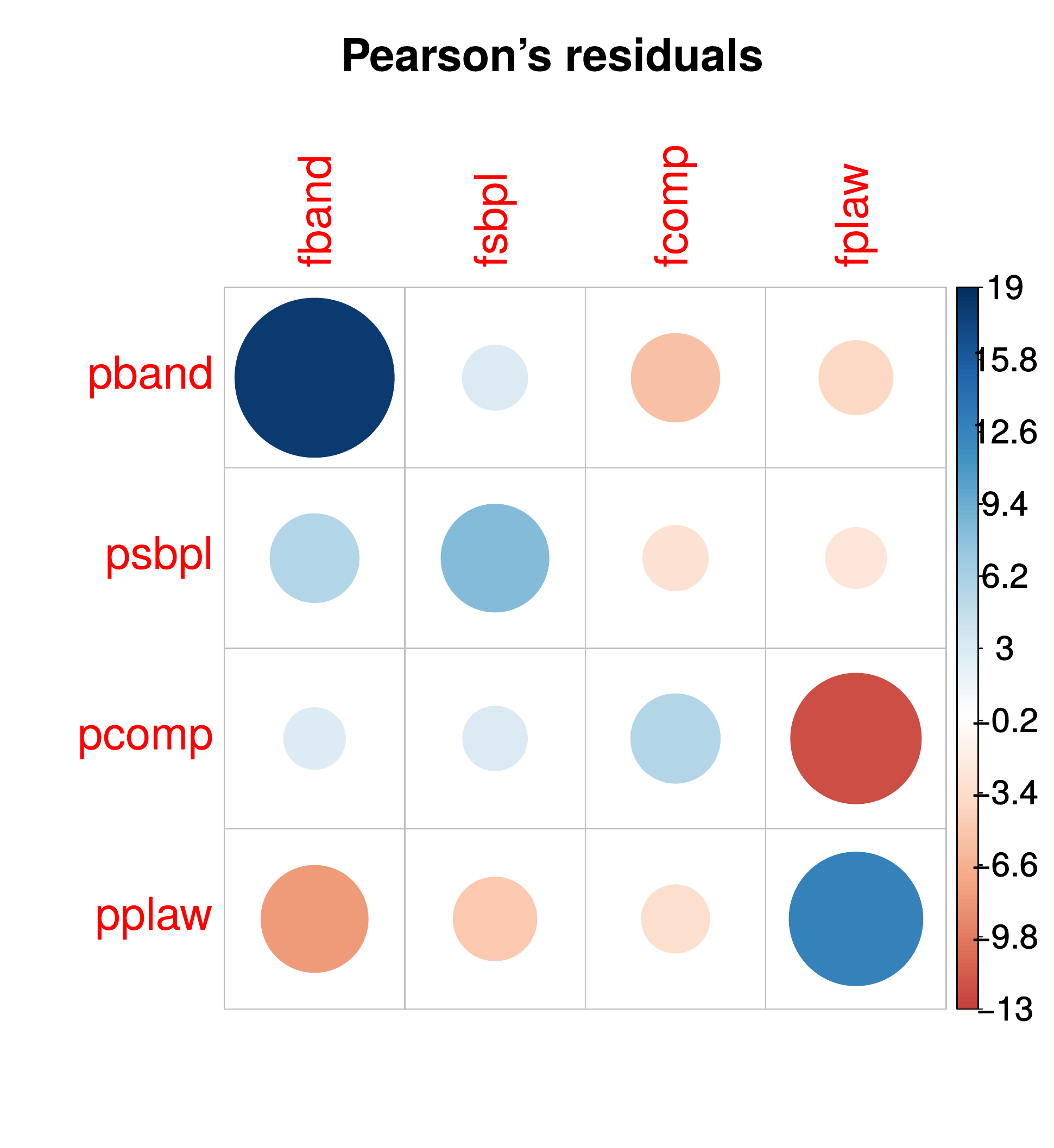}
  \caption{Contribution of the $s^2_{ij}$ squared residuals to the total
$\chi^2_{df}$ value in  he left side of Equation~(\ref{chi2})
(left panel) and $s_{ij}$ Pearson's residuals (right panel). The scale in the left panel is in percentage. The blue color in the right panel means surplus and the red a deficit with respect to the expected value assuming no relationship between the peak flux and fluence spectra. Both panels  show clearly that the strongest contribution to the departure from the homogeneous (random) case is given by the frequencies of the power law and Band spectra.}\label{ch2res}
\end{figure}

Computing the $s^2_{ij}/\chi^2_{df}$ ratios we get the fractional strengths of the contribution of the individual cell of Table~\ref{contab} to the left side of Equation~(\ref{chi2}). It is shown in Fig.~\ref{ch2res} along with the $s_{ij}$ normalized (Pearson's) residuals.

One can infer from Fig.~\ref{ch2res} that the power law and Band spectra gives the strongest deviation from the homogeneity of Table~\ref{contab}. This means there is a tight correlation between getting a power law from the peak flux and fluence, respectively. The same is hold for the Band spectra. On the contrary, there is a little chance of getting a power law from the fluence assuming a Comptonized spectrum was obtained from the peak flux. The large red spot at the right bottom corner of the colored Table in Fig.~\ref{ch2res} can be explained in this way.

\subsection{Correspondence analysis (CA)}
\label{CAanal}

The main objectives of CA \citep{CA2,CA3} are to transform a dataset into two set of factor scores (rows and columns) that give the best representation of the similarity structure of the rows and columns of the table.

\subsubsection{Mathematical Background}

The $s_{ij}$ Pearson's residual matrix in Equation~(\ref{stvar}) expresses the deviation from the homogeneity of the contingency table. This matrix can be factorized by using the singular value decomposition (SVD) into the following form:

\begin{equation}\label{SVD}
   s_{ij}=\sum\limits^{nd}_{k=1}\sum\limits^{nd}_{l=1}u_{ik}\Delta_{kl}v_{lj}^T
   \phantom{@@@}
\end{equation}

\noindent where $u_{ik}$ relates to the rows and $v_{jl}$ to the columns of $s_{ij}$. $\Delta_{kl}$ is a diagonal matrix containing the singular values in the diagonal (the $T$ exponent denotes transpose and $nd=min(nr,nc)-1=3$ in our  case).

Denoting with $\delta_k$ the singular values obtained in the SVD their sum of squares relates to the left side of Equation~(\ref{chi2}).

\begin{equation}\label{ssqrs}
     \chi^2_{df}=N\sum \limits^{nd}_{k=1}\delta_k^2
\end{equation}

\noindent $N$ is the sample size (in our case 1852). The left side of the above Equation expresses the level of deviation from the homogeneity of Table~\ref{contab}. The following equality expresses the relative importance of $\delta_k$ in Equation
(\ref{ssqrs})

\begin{equation}\label{rimp}
   w_k=\frac{\delta_k^2}{\sum\limits^{nd}_{k=1}\delta_k^2}
\end{equation}

The columns of the $u_{ik}$ and $v_{jl}$ form an orthogonal base, respectively. Considering the rows (columns) of $s_{ij}$ as variables they can be represented as points in the coordinate systems defined by these orthogonal vectors obtained from $u_{ik}$ and $v_{jl}$. To represent the row (column profiles, respectively) in these orthogonal systems one can define coordinates for the row profiles as

\begin{equation}\label{fcoord}
    f_{il}=r_i^{-1}\sum\limits^{nd}_{i=1}u_{ik}\Delta_{kl}
\end{equation}

\noindent and for the column profiles as

\begin{equation}\label{gcoord}
    g_{jl}=c_j^{-1}\sum\limits^{nd}_{k=1}v_{jk}\Delta_{kl}.
\end{equation}

\noindent Using these coordinates one can display the similarities (dissimilarities) between the different profiles graphically.

In a CA map when two row (respectively, column) points are close to each other means that these points have similar profiles. The position of a row (respectively, column) point is determined from its barycenter on the column (respectively, row), therefore the proximity between a row point and one column point cannot be interpreted directly.

\subsubsection{CA of the GBM Data}

There are several libraries in R containing a procedure for the CA. We used the CA available in the $FactoMineR$ library.

Performing CA on Table~\ref{contab} we obtained the coordinates of Equation~\ref{fcoord}~and~\ref{gcoord} presented in Table~\ref{utab} and Table~\ref{vtab}. The zero value (origo) of the Dimensions in the Tables refers to the average profile. The greater the deviation of the Dimensions from the zero value, the greater the departure of a profile (row or column) from the average one.

\begin{table}
\centering
\begin{tabular}{rrrr}
  \hline
 Type of spectra & Dim 1 & Dim 2 & Dim 3 \\
  \hline
  Band & 2.11 & 1.25 & -0.31 \\ 
  Sb. power law & 1.58 & 0.51 & 1.13 \\ 
  Comptonized & 0.35 & -0.34 & -0.02 \\ 
  Power law & -0.49 & 0.19 & 0.01 \\ 
   \hline
\end{tabular}
\caption{Results of CA for the peak flux spectra. The columns of Dim 1-3 contain the coordinate values computed according to Equation~(\ref{fcoord}). The coordinates of Dim 1 give the greatest discrimination power. The greatest difference is between the Band and power law spectral type in accordance to Sect.~\ref{simcr}.} \label{utab}
\end{table}

\begin{table}
\centering
\begin{tabular}{rrrr}
  \hline
 Type of spectra & Dim 1 & Dim 2 & Dim 3 \\
  \hline
  Band & 1.46 & 0.59 & -0.17 \\ 
  Ssb. power law & 0.86 & -0.04 & 0.60 \\ 
  Comptonized & -0.01 & -0.27 & -0.04 \\ 
  Power law & -0.73 & 0.46 & 0.03 \\ 
   \hline
\end{tabular}
\caption{Results of CA for the fluence spectra. The values in columns Dim 1-3 are the coordinates of column profiles obtained from applying Equation~(\ref{gcoord}) to the GBM data. Similarly to Table~\ref{utab} the greatest difference is between the profiles of power law and Band spectra.} \label{vtab} \end{table}

\begin{figure}
\centering
  \includegraphics[width=6cm]{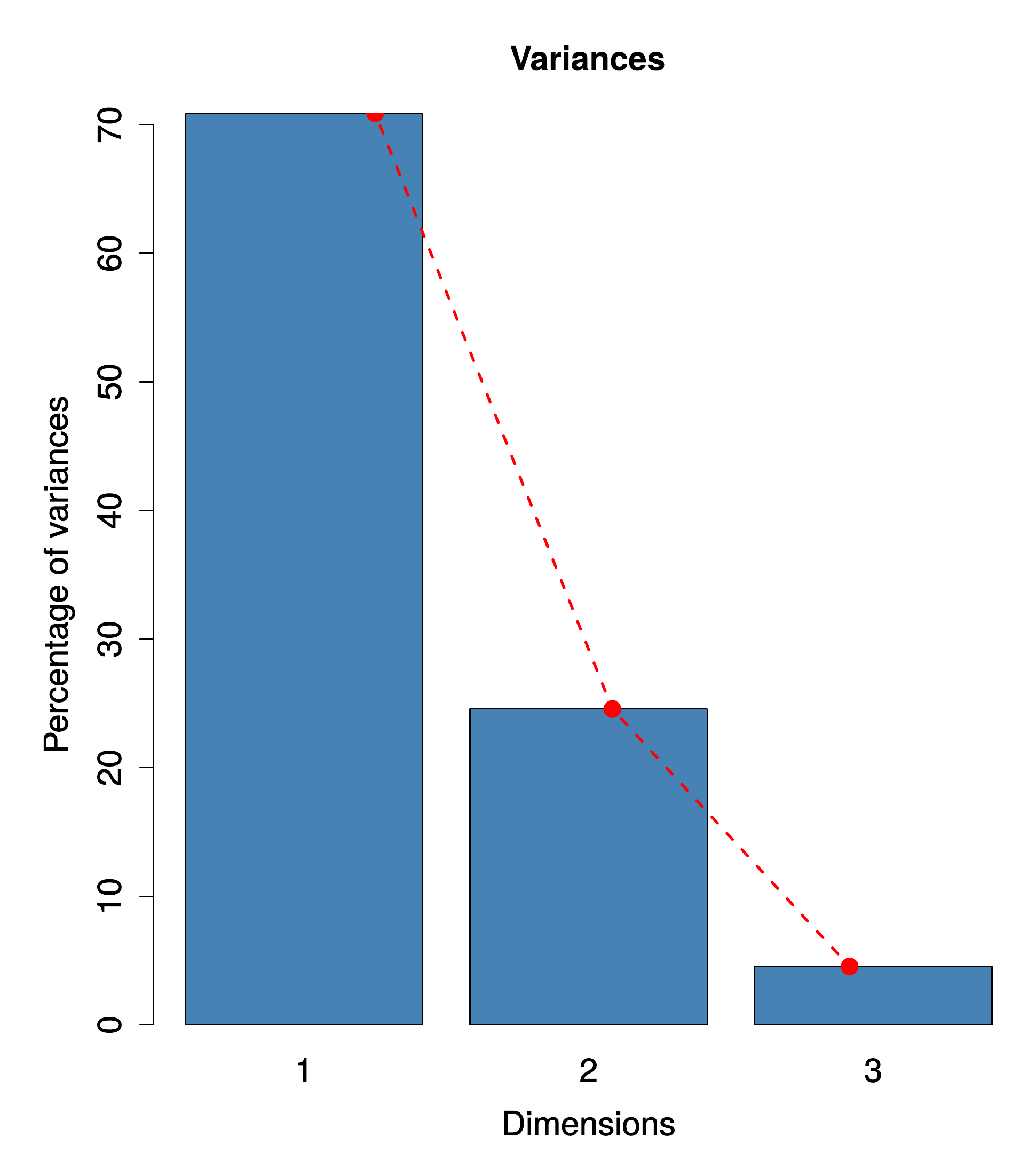}
\caption{Relative importance ($w_k$) of the Dimensions associated to the $\delta_k^2$  values (variances). We neglected the variance associated to Dimension 3.}\label{var}
\end{figure}

\begin{figure}
\centering
  \includegraphics[width=8cm]{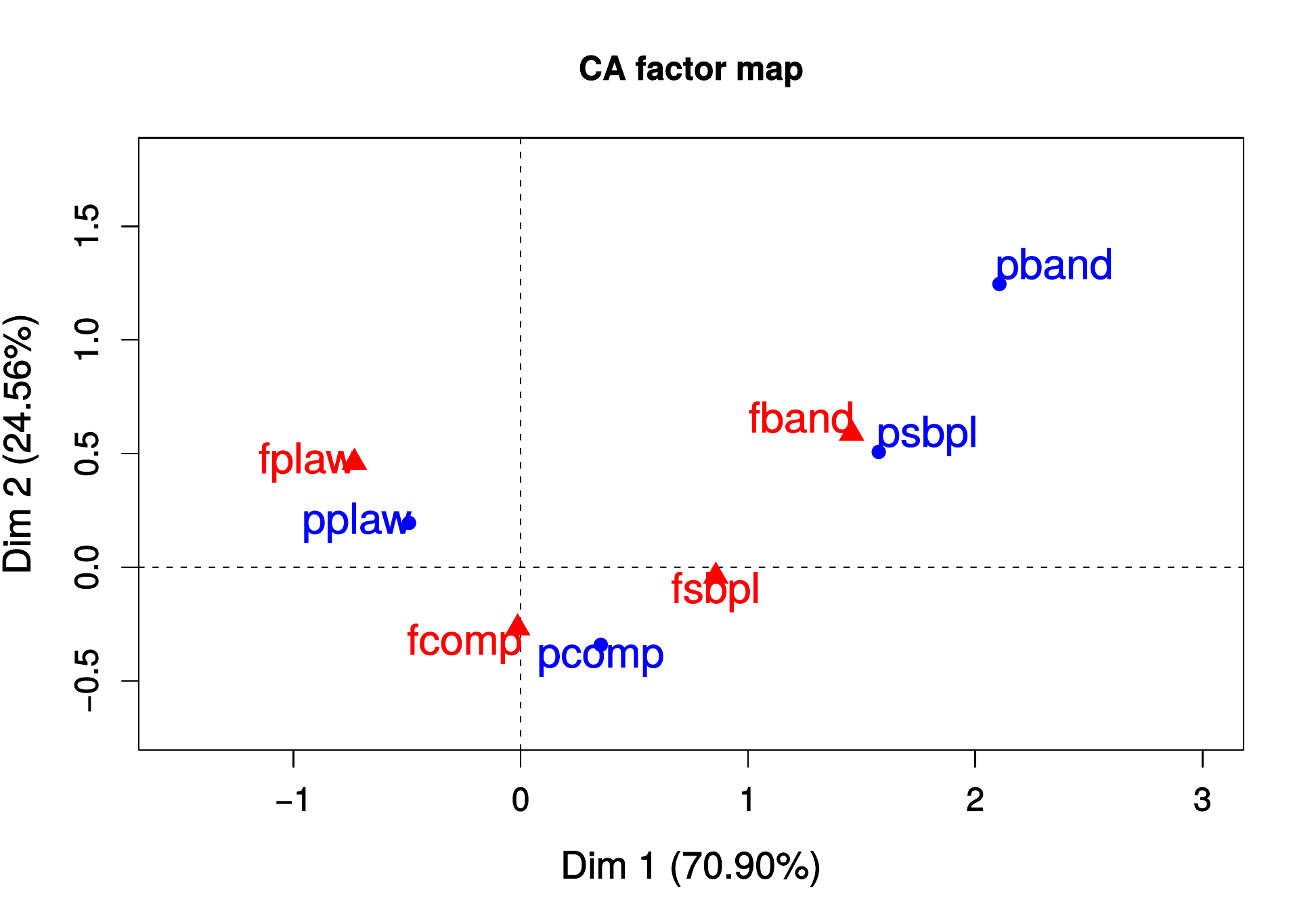}
 \caption{Graphical display of the CA results. The percentages in brackets give the $w_1$ and $w_2$ values obtained from Equation~(\ref{rimp}) for the associated singular values in Equation~(\ref{SVD}). Note that the greatest difference is between the power law and the Band spectra, both in the peak flux and in the fluence.}\label{CAmap}
\end{figure}

According to Equation~(\ref{rimp}) we computed the $w_k$ weights associated  with the $\delta_k$ singular values and displayed them in Fig.~\ref{var}. One can infer from this Figure that $\delta_1^2$, the square of the first singular value explains more than 70 \% of the sum of squares in the right side of Equation~(\ref{ssqrs}). Correspondingly, the associated $Dim \,1$, computed from Equations (\ref{fcoord})~and~(\ref{gcoord}) and presented in Table~\ref{utab}~and~\ref{vtab}, give the greatest discriminant power (distance) between the row and column profiles, respectively (see also Fig.~\ref{CAmap}).

Using $Dim$ 1 and 2 we displayed in Fig.~\ref{CAmap} the results of the CA. All dots in this Figure represent a profile belonging to one of the spectral types concerning the peak flux (blue color) and the fluence (red color), respectively. The origo of this image represent the average profile. The deviation of a given dot from the origo measures the departure of the corresponding profile from the average one.

The difference between the dots characterize quantitatively the difference between the corresponding profiles. It refers, however, only to the profiles belonging to the peak flux or to the fluence, separately. Namely, the dots in Fig.~\ref{CAmap}  refer to their own average Profile which are different at the rows and at the columns. The direct comparison of the dots belonging to peak flux with those to fluence is not meaningful.

In the positions of the dots in Fig.~\ref{CAmap} we recovered the rule obtained in Sect.~\ref{simcr}: the spectra can be ordered into a series starting at the power law and ended at the Band spectrum. The two end points of this series have the greatest dissimilarity in their profiles.

Summing up the results obtained in Sect.~\ref{ctanal}~and~\ref{CAanal} we concluded: there is an extremely significant correspondence between the spectral types  belonging to the peak flux and to the fluence. This correspondence, however is not one for one.

The spectra can be ordered into a series starting at the power law and ended at the Band spectrum. Proceeding from a spectrum of the peak flux to that of the fluence the type can be changed towards the direction of the Band but backwards only with a very low probability.

\subsection{Discriminant Analysis}

Discriminant analysis aims to recognize difference between groups in the multivariate parameter space, orders membership probabilities to the cases and one may use this scheme for classifying additional ones not having assigned group memberships. We use this technique to look for differences in the distributions  of GRBs, in the parameter space defined by the variables listed in the Fermi GBM catalog.

There are several approaches to solve this problem. We used the linear discriminant analysis (LDA) originally proposed by \citet{LDA1}; see \citet{LDA2} for a review. In the following we summarize its basic mathematics.

\subsubsection{Mathematical background}

Let we have a set of $p$ measured variables on $n$ cases which are assigned to one of the $k$ classes ($k=4$ in our case). We look for linear combination of the $x_1,x_2, \ldots, x_p$ variables which give maximal separation between the groups of the cases. It means we are looking for the variables

\begin{equation} \label{ydis}
y=n_1x_1+n_2x_2+ \ldots +n_px_p \phantom{@@@} \mbox{where} \phantom{@@@}
\sum^p_{j=1}n^2_j=1
\end{equation}

\noindent with a suitable chosen $n_1, n_2, \ldots ,n_p$ coefficients ensuring a maximal separation between the classes.

To get the most discriminating direction in the parameter space, defined by the measured variables, one computes the variance of the $y$ variable in Equation~(\ref{ydis}) within the groups.

\begin{equation}\label{within}
    \sigma^2_w =\frac{\sum \limits^k_{l=1} \sum
    \limits^n_{i=1}\delta_{il}(y-\mu_l)^2}{n-1}
\end{equation}

\noindent where $\mu_l$ is the mean value of $y$ within the groups and $\delta_{il}$ is the Kronecker delta ($\delta_{il}=1$ if $i=l$ and 0 otherwise).

To characterize the differences between the groups along the $y$ variable one compute the variance of the $\mu_l$ group means

\begin{equation}\label{ybetw}
    \sigma^2_b = \frac{\sum \limits^k_{l=1}(\mu_l-\mu)^2}{k-1}.
\end{equation}

\noindent In the above Equation $\mu$ is the sample mean of the $y$ variable.

LDA looks for a $n_1, n_2, \ldots ,n_p$ direction in the parameter space of $p$ dimension along which the ratio of the variances of $y$ between and within the groups

\begin{equation}\label{rbw}
    \lambda = \frac{\sigma^2_b}{\sigma^2_w}
\end{equation}

\noindent is maximum. In LDA $\lambda$ can be obtained from an eigen value problem and the best discriminating direction is an eigen vector belonging to $\lambda$. One can continue this procedure in the subspace perpendicular to this eigen vector and get a further eigen value and vector. Altogether k-1 discriminating directions can be obtained.

It is an important question whether discriminating  the groups along these directions is really significant. To test the significance one computes the following variable, the Wilk's lambda

\begin{equation}\label{wlam}
    \Lambda = \prod \limits^{k-1}_{l=1}\frac{1}{1+\lambda_l}.
\end{equation}

\noindent Assuming the null hypothesis, i.e. the differences between the groups are only accidental $\Lambda$ can be related to a $\chi^2$ distribution

\begin{equation}\label{lsig}
    \chi^2_{df} = -(n-\frac{p+k}{2}-1)ln\Lambda
\end{equation}

\noindent where $df=p(k-1)$ is the degree of freedom of the $\chi^2$ variable \citep{LDA3}. Omitting the first eigen value and vector one can continue this procedure in the subspace perpendicular to the best discriminating direction. Of course, $\lambda_1$ has to be omitted in Equation~(\ref{wlam}) and $p$ has to be replaced with $p-1$ in Equation~(\ref{lsig}).

LDA is usually among the major ingredients of the professional statistical software packages. We used the {\tt lda()} procedure available in the $MASS$ library \citep{MASS} of the R statistical package.

\subsubsection{LDA of the GBM Data}
\label{ldgb}

Making use of LDA we examined 2069 GRBs of the the Fermi GBM catalog. For the input variables we used the model independent physical parameters mentioned above ($T90$, $T50$, fluence, batse fluence, 64ms/256ms/1024ms flux, batse 64ms/256ms/1024ms flux, and the categorical variables the spectral types of both peak flux and fluence). 

The LDA resulted in three discriminant functions (LD1,~LD2,~LD3). The mean values of the discriminant functions within the different spectral types and significance obtained in this way are summarized in Table~\ref{tab:lda_full}. The output of this procedure also gave information on the significance of the discrimination between the spectral types, associated with the three discriminant functions. The level of significance is based on the probability of the value of the Wilks' lambda, defined in Equation~(\ref{wlam}).

Table~\ref{tab:lda_full} demonstrates that the greatest discriminated distances appears to be between the Band and power law types, with the remaining two types are between them. It is true for both of the peak flux and the fluence spectra. One sees, however, that the level of discrimination is higher in case of the peak flux than the fluence spectra. 

\begin{figure}
\centering
\includegraphics[width=4cm]{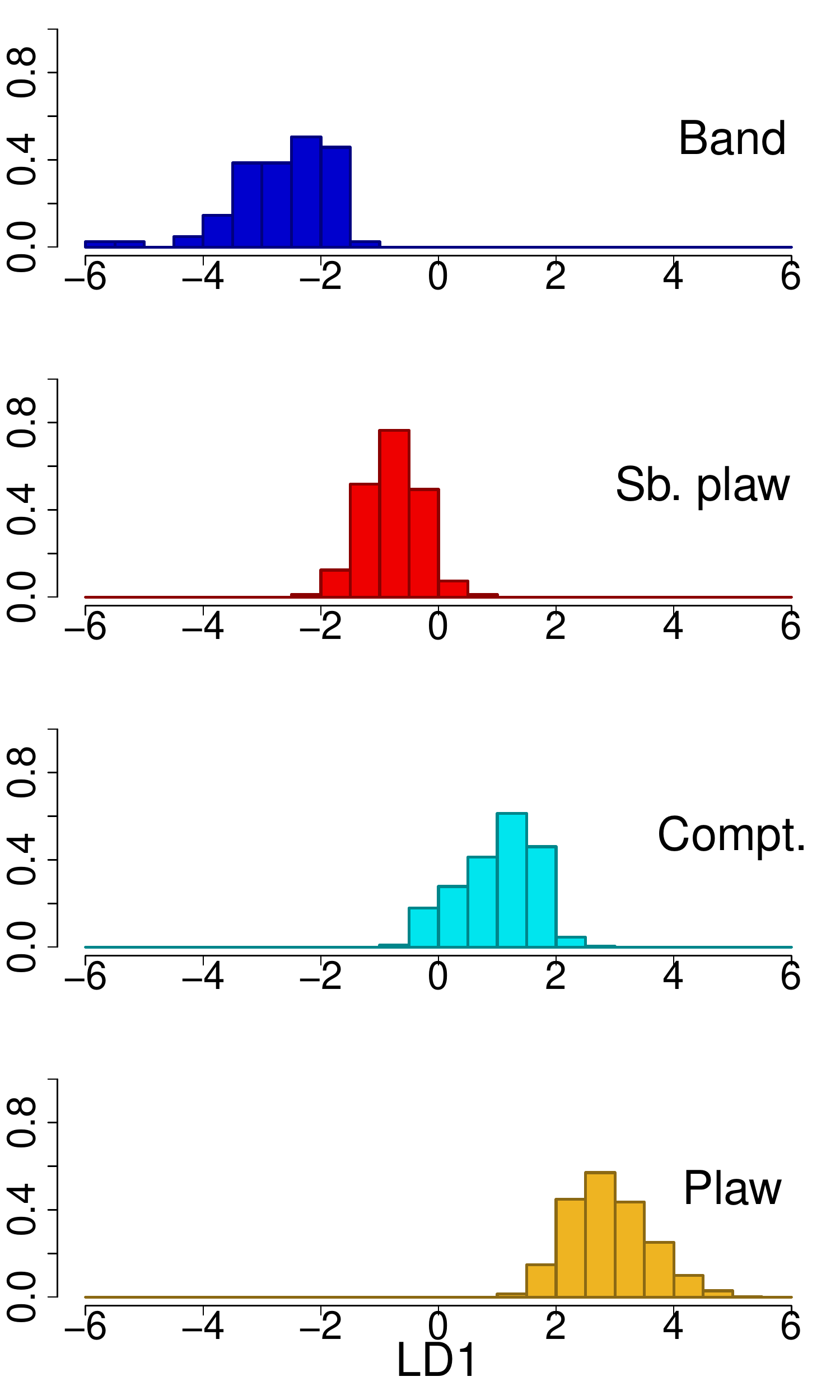}
\includegraphics[width=4cm]{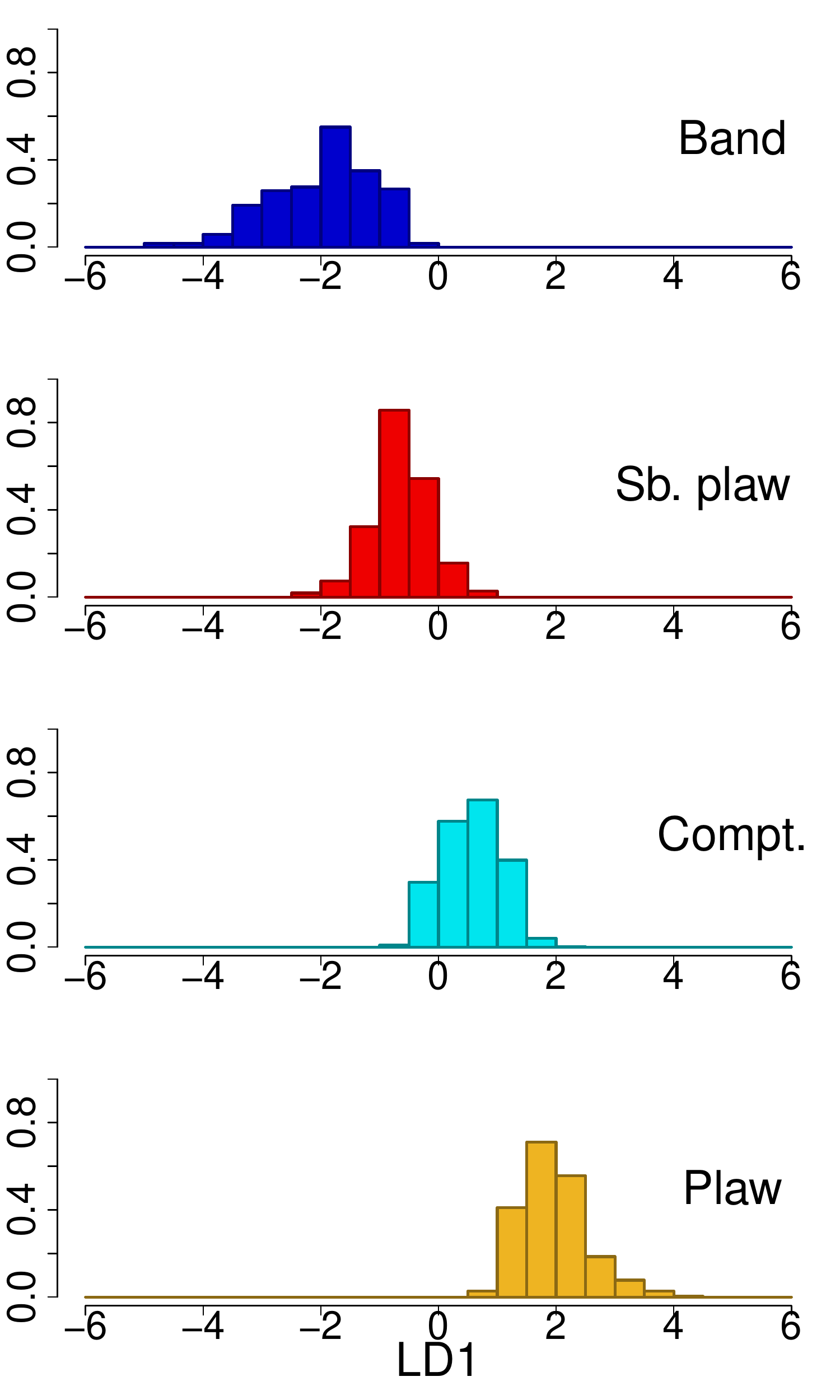}
\caption{\label{fig:lda_full} The distribution of the first discriminant function between the spectral classes and types, on the left side the 'peak flux' types and on the right the 'fluence' type models. We found that the difference between the classes are bigger in the 'peak flux' models but both (\textit{pflx} and \textit{flnc}) significances are very high.}
\end{figure}

This results obtained for the peak flux and fluence spectra are also shown in Fig.~\ref{fig:lda_full}. One can see that in both cases the first linear discriminant function (LD1) separates the different spectral types well. Comparing the two plots one can see that fitting the model spectra on the peak flux is much more efficient (see the Table~\ref{tab:lda_full} for the significance values). The results are in line with the results of the correspondence analysis, so that the power law and the Band spectral types are the most distant from each other. The LDA separates the power law and the Comptonized types well (in case of peak flux classification), however, the Band and the smoothly broken power law merge into each other.

\begin{table}
\centering
\begin{tabular}{r|ccc}
  \hline
\textbf{Peak flux classes}  & LD1 & LD2 & LD3 \\ 
  \hline
Band & -2.363 & 0.111 & -0.130 \\ 
  Sb. plaw & -1.215 & 0.072 & 0.176 \\ 
  Comptonized & 0.744 & -0.414 & -0.019 \\ 
  Power law & 2.835 & 0.231 & -0.027 \\ 
  \hline
  Sign. level & $<2.2\cdot 10^{-16}$ & 0.311 & 0.99 \\ 
   \hline
\multicolumn{4}{c}{} \\
  \hline
\textbf{Fluence classes} & LD1 & LD2 & LD3 \\ 
  \hline
Band & -1.676 & 0.242 & -0.061 \\ 
  Sb. plaw & -0.819 & -0.321 & -0.070 \\ 
  Comptonized & 0.524 & 0.010 & 0.289 \\ 
  Power law & 1.970 & 0.069 & -0.158 \\ 
  \hline
  Sign. level & $<2.2\cdot 10^{-16}$ & $2.1\cdot 10^{-4}$ & 0.625 \\ 
   \hline
\end{tabular}
\caption{\label{tab:lda_full} Mean of the discriminant functions between the spectral classes. The table shows typical values of discriminant functions and their significance levels. To get the significance we used the $\chi^2$-test (see Equation~\ref{lsig}). The first LDs are very highly significant in both spectral types. At the second functions only the \textit{flnc} is relevant and we present the cause of this in Sect.~\ref{sec:lda_power}.}
\end{table}

\begin{figure}
\centering
\includegraphics[width=8.25cm, trim=0 0 0 1.5cm]{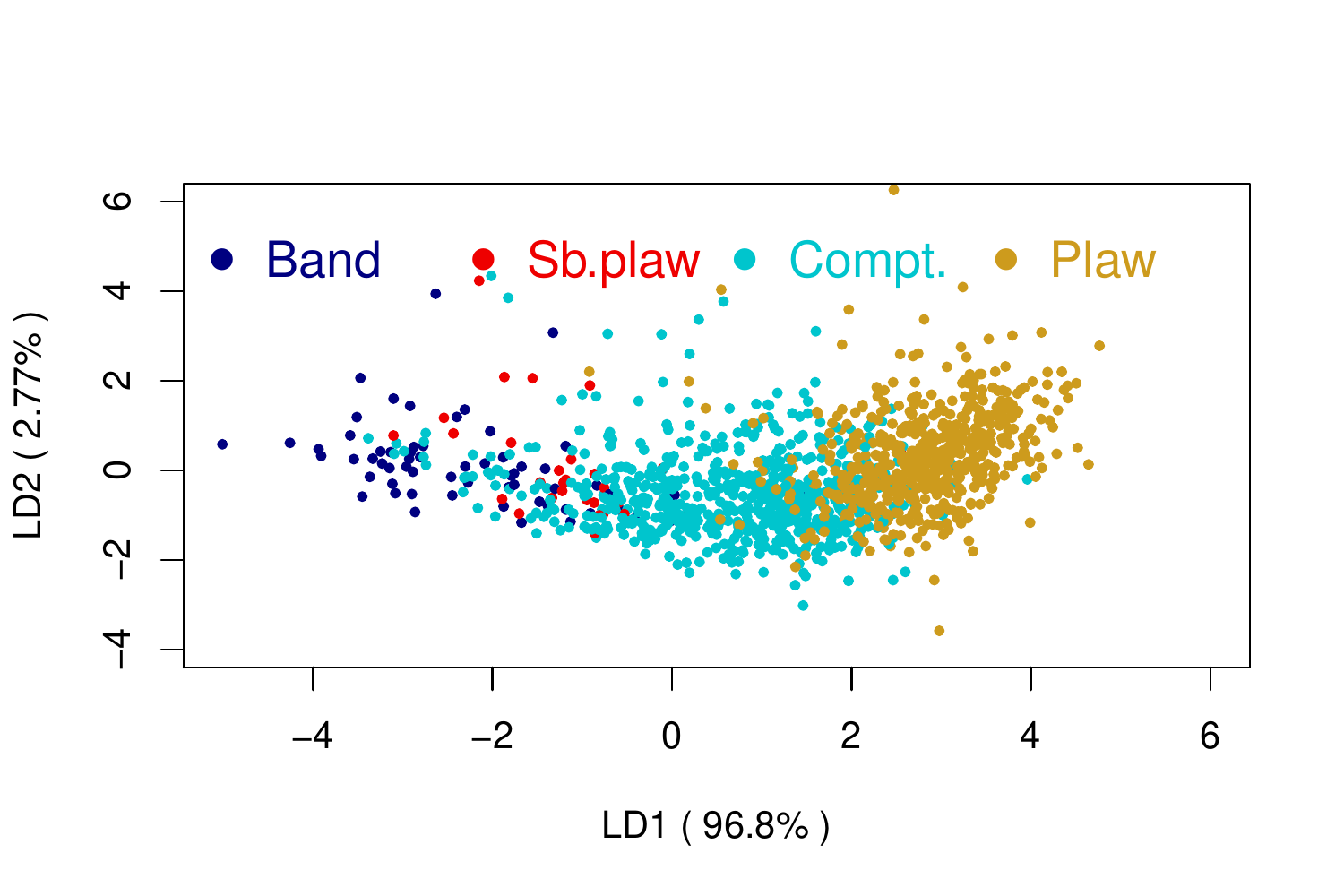}
\includegraphics[width=8.25cm, trim=0 0 0 1.5cm]{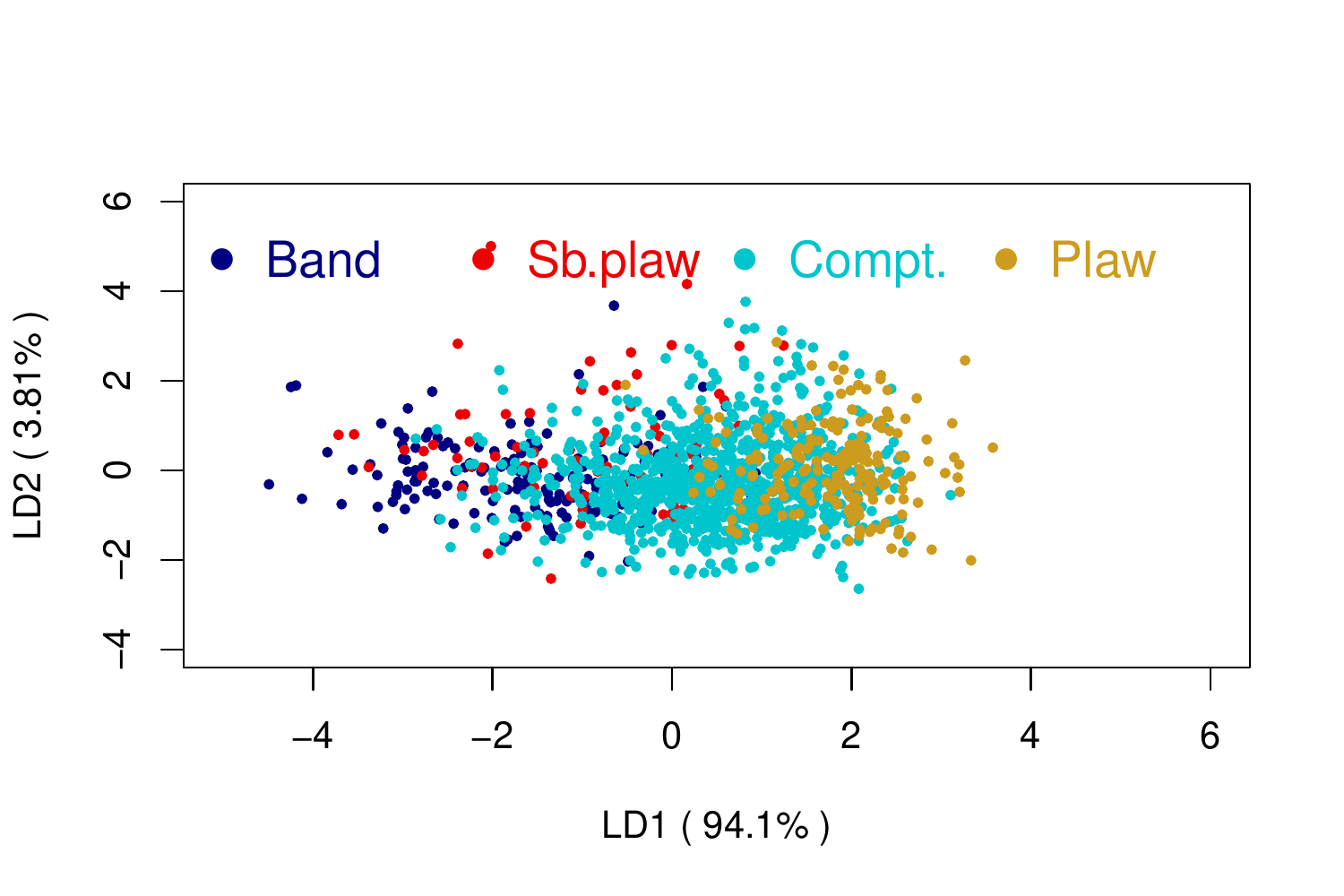}
\caption{\label{fig:lda_good} Plot of the two main discriminators on both spectral types from the 'GOOD sample' of GRBs. The criteria of this sample was published by \citet{fermi1}. The top figure shows the \textit{pflx} models and the bottom the \textit{flnc}. We found that peak flux models are more segregated with the first discriminant function.}
\end{figure}

Until this point we used the whole complete sample without taking into account the different uncertainties of the data given in the catalogue. \citet{fermi1}, however, defines a subsample of 'GOOD' taking into account only those objects having uncertainties below a certain limit. We repeated LDA with the data fulfilling these criteria. Performing LDA in this case resulted in improved significances and the peak flux spectra of the LD2 became significant as well (sign: $8.5\cdot 10^{-4}$). The typical values of classes on every function and the significance levels can be seen in Fig.~\ref{fig:lda_good} and Table~\ref{tab:lda_good}. 

\begin{table}
\centering
\begin{tabular}{r|ccc}
  \hline
\textbf{Peak flux classes} & LD1 & LD2 & LD3 \\ 
  \hline
Band & -2.345 & 0.145 & -0.156 \\ 
  Sb. plaw & -1.141 & 0.106 & 0.214 \\ 
  Comptonized & 0.693 & -0.554 & -0.021 \\ 
  Power law & 2.794 & 0.303 & -0.038 \\ 
  \hline
  Sign. level & $<2.2\cdot 10^{-16}$ & $8.5\cdot 10^{-4}$ & 0.995 \\ 
   \hline
\multicolumn{4}{c}{} \\
  \hline
\textbf{Fluence classes} & LD1 & LD2 & LD3 \\ 
  \hline
Band & -1.542 & -0.195 & -0.183 \\ 
  Sb. plaw & -0.756 & 0.388 & 0.109 \\ 
  Comptonized & 0.495 & -0.260 & 0.253 \\ 
  Power law & 1.803 & 0.067 & -0.180 \\ 
  \hline
  Sign. level & $<2.2\cdot 10^{-16}$ & $4.1\cdot 10^{-6}$ & 0.069 \\ 
   \hline
\end{tabular}
\caption{\label{tab:lda_good} Mean of the discriminant functions between the spectral classes and their significances from the 'GOOD sample'. The significances are strongly increased and both second discriminant functions became important separators.}
\end{table}

\subsubsection{LDA of GRBs jointly detected with the Swift satellite}

Both the Swift and the Fermi satellites opened a new area in studying GRBs. There are significant differences in the covered energy  range, the technical layout and observing strategy of these instruments. As a consequence there are differences in the GRB populations detected by these satellites. It is interesting, however, to derive the overlap between the populations of the Swift and Fermi detected GRBs and to look for differences from those not observed jointly. 

\begin{figure}
\includegraphics[width=8.25cm]{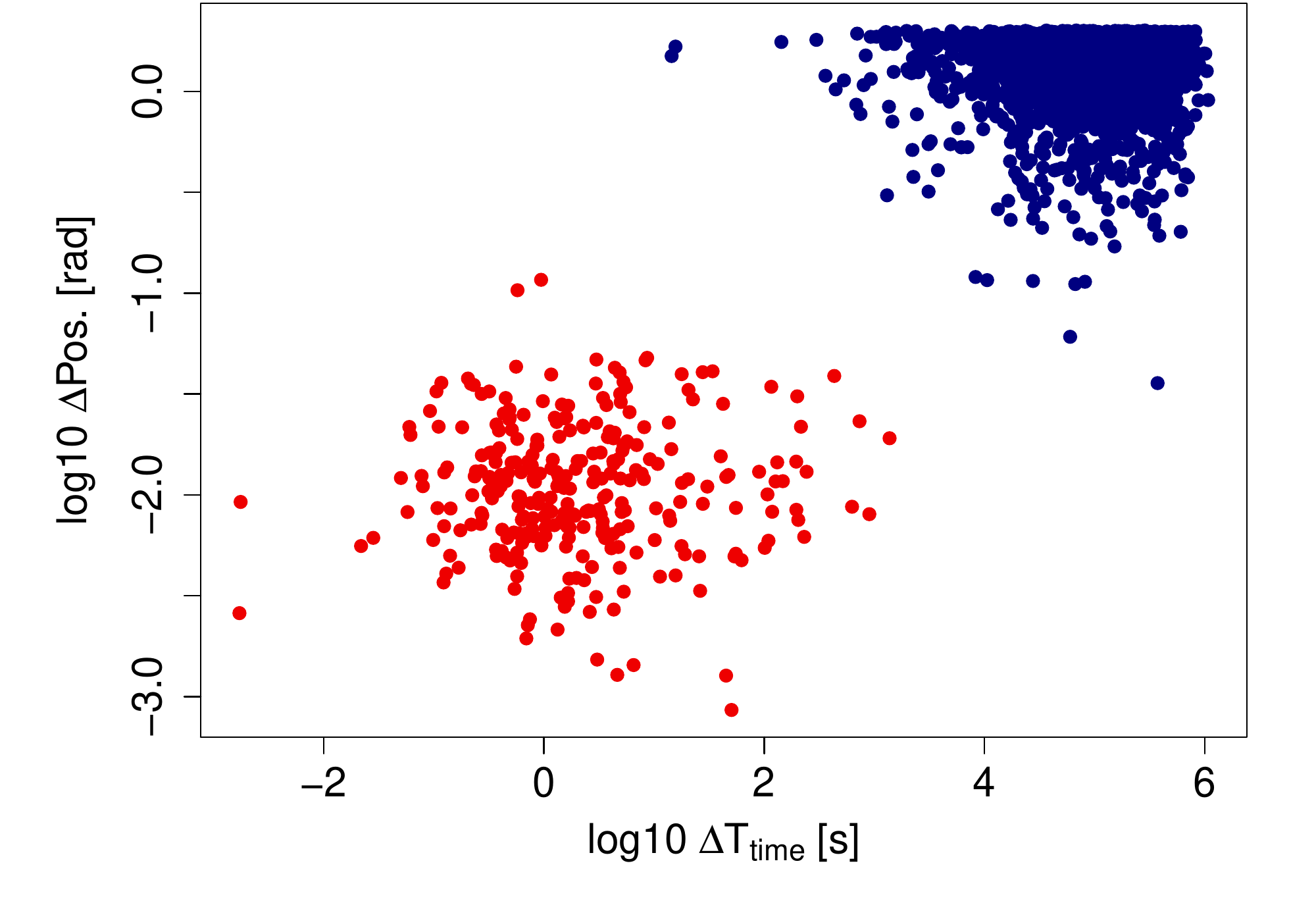}
\caption{\label{fig: fermi_swift} Grouping the matched Swift-Fermi GRBs. We used Model-Based Clustering in the R package to select the Swift-Fermi pairs and we found a well separated group (red dots) where the time difference is smaller than $\sim$30 minutes and the coordinates are closer than $\sim$10 degrees.}
\end{figure}

To get the jointly detected GRBs we compared the objects presented in the GBM catalog with those recorded by Swift. To find the jointly detected objects we looked for the closest neighbor  in the other catalogue, in position and time. For the Swift BAT GRB objects the trigger time and position data were taken from the NASA GRB Table database\footnote{\url{https://swift.gsfc.nasa.gov/archive/grb_table/}}. We calculated the elapsed time from a reference time (2000-01-01T00:00) for both the Fermi and the Swift GRBs. To find the Fermi-Swift data pairs of the shortest time and spatial differences we used the {\tt knn()} procedure from the $class$ library of the R package \citep{MASS}. Fig.~\ref{fig: fermi_swift} shows our results. We found a well separated group with the Model-Based Clustering procedure {\tt Mclust()}) from the $mclust$ package \citep{mclust1,mclust2}.

This group of GRBs is clearly defined by the time difference of less than 30 minutes and the coordinate difference is less than 10 degrees. 

It is known that the Swift telescope has a couple of arcminutes error in measuring positions but in case of Fermi it can go up to some degrees. Within the boundaries set by the parameters above we found 300 Fermi-Swift GRB pairs. After checking the data manually we found 8 incorrect associations in which there were 3 Swift GRBs matched with 2 Fermi GRBs each and there were 2 false data points, as a not real Swift detection. (For further details of the LDA analysis of this sample see Sect.~\ref{sec:fermi_swift_disc}.)

According to their physical character of the model independent physical variables we used in LDA can be ordered into three major groups: duration, peak flux and fluence. We selected three representative variables of each types and displayed their mean values in Table~\ref{tab:kruskal}, within all spectral groups (for more details see Sect.~\ref{sec:lda_power}).

We performed Kruskal-Wallis tests for all the selected representative variables to test whether there are differences between spectral groups in the given variable. For this purpose we used the {\tt kruskal.test()} procedure from R $stats$ package \citep{R}. Table~\ref{tab:kruskal} shows the significance values obtained from the Kruskal-Wallis rank sum test. It shows that the duration is not a  discriminant parameter between the spectral types. We can also see that the band and smoothly broken power law don't differ according to these parameters. We found that there are only three significantly different spectral classes, the Band, the Comptonized, and the power law groups for the sample of the jointly detected GRBs.

\subsubsection{LDA of GRBs having measured redshift}
\label{sec:fermi_swift}

We selected the Fermi-Swift GRB pairs with available redshift data. We used here the the GRB tables made by Jochen Greiner\footnote{\url{http://www.mpe.mpg.de/~jcg/grbgen.html}} containing most of the redshift data for GRBs. We found 74 Fermi-Swift GRBs which have spectroscopic redshift. Using the redshift we calculated the emitted peak-flux (Equation~\ref{eq:co_moving_flux}) and the fluence (Equation~\ref{eq:co_moving_fluence}) in the co-moving system from the observed peak-flux and fluence. The calculation was based on the Equation~(2) from \citet{meszaros_cosmo_calc} with the {\tt D.L()} procedure from the 'cosmoFns' R package \citep{cosmoFns} for the luminosity distance calculation. 

\begin{equation}
\mbox{Flux(z)}=\frac{\mbox{Flux}_{\mbox{obs}} 4\pi d_L^2}{1+z}
\label{eq:co_moving_flux}
\end{equation}
\begin{equation}
\mbox{Fluence(z)}=\frac{\mbox{Fluence}_{\mbox{obs}} 4\pi d_L^2}{1+z},
\label{eq:co_moving_fluence}
\end{equation}

where z is the redshift of the GRB, $d_L$ is the luminosity distance of the source and $\mbox{Flux}/\mbox{Fluence}_{\mbox{obs}}$ are the observed Flux/Fluence.

In addition, we transform the $T90$ and the $T50$ values into the co-moving systems with the Equations~(\ref{eq:co_moving_T90}).

\begin{equation}
T90/50(z)=\frac{T90/50_{\mbox{obs}}}{1+z}
\label{eq:co_moving_T90}
\end{equation}
With these new parameters we repeated the LDA with both the fluence and the peak-flux spectral classification.

\begin{table}
\centering
\begin{tabular}{r|rrr}
  \hline
\textbf{Peak flux classes} & LD1 & LD2 & LD3 \\ 
  \hline
Band & 2.560 & 0.944 & 0.217 \\ 
  Sb. plaw & 0.855 & -1.583 & 0.065 \\ 
  Comptonized & -0.613 & 0.332 & -0.641 \\ 
  Power law & -2.802 & 0.307 & 0.359 \\ 
  \hline
  Sign. level & $7.6\cdot 10^{-4}$ & 0.987 & 0.999 \\ 
   \hline 
\multicolumn{4}{c}{} \\
  \hline
\textbf{Fluence classes} & LD1 & LD2 & LD3 \\ 
  \hline
Band & -2.114 & 0.589 & 0.108 \\ 
  Sb. plaw & -0.509 & -0.937 & 0.213 \\ 
  Comptonized & 0.525 & -0.025 & -0.643 \\ 
  Power law & 2.098 & 0.373 & 0.321 \\ 
  \hline
  Sign. level & $6.3\cdot 10^{-4}$ & 0.997 & 0.999 \\ 
   \hline
\end{tabular}
\caption{\label{tab:lda_redshift} The comoving discriminant functions. We calculated the LDA functions after the co-moving transformation and this table shows mean values of the discriminant functions. It is clear that significances are strongly reduced with respect to the observers frame.}
\end{table}

We compared the redshift distribution of the spectral groups, and did not find differences between the distributions of both \textit{pflx} and \textit{flnc} types, within the limits of statistical inference. We performed LDA with the variables transformed into the co-moving system. We found the discriminant significance decreased a little but the first LD remained significant. From Table~\ref{tab:lda_redshift} the level of significance of LD1 are $7.6\cdot 10^{-4}$ (from the peak flux type) and $6.3\cdot 10^{-4}$ (from the fluence type).

We should mention, however, that the categorical variables referring to the model spectra were fitted to the spectral distribution of the observed photons. Consequently, they cannot be transformed onto the comoving frame. The correct procedure would be to transform each observed photons into the comoving frame and repeat the procedure to get the best fitting model. One should mention, however, that this procedure is justified only for those photons coming directly from the GRB itself. Since the noise level is also detected one can correct it statistically.

Although we get the true spectral distribution of the photons in this transformation, their total number is identical to the observed one. Since the redshift of these GRBs is known the true number of photons can be estimated. To proceed in this way, however, is beyond the scope of this paper.

\section{Discussion}
\label{disc}

\subsection{Discriminant power of the observed variables}

\subsubsection{Discriminant power of peak flux and fluence}
\label{sec:lda_power}

We calculated which physical parameters contribute significantly to the discriminant functions. The strength of this contribution can be characterized by the correlations between the input (physical) and the computed discriminant variables, by the structure matrix in Table~\ref{tab:lda_function}. We can see that in case of peak flux classification the most important parameter is the flux, while in the fluence classification the fluence is the most important in the first discriminant function. 

Although in both cases the LD1, LD2 and LD3 discriminant functions can not be identified with a single physical variable, they are produced by the combination of several physical parameters. It is also true for LD1, although it has a very high level of correlations with the input variables.

\begin{table}
\centering
\begin{tabular}{r|rrr}
  \hline
 \textbf{Peak flux classes} & LD1 & LD2 & LD3 \\ 
  \hline 
  t50 & -0.146 & -0.013 & \textbf{0.431} \\ 
  t90 & -0.053 & 0.042 & \textbf{0.213} \\ 
  fluence & \textbf{0.805} & \textbf{0.151} & \textbf{0.262} \\ 
  fluence\_batse & \textbf{0.802} & \textbf{0.181} & \textbf{0.321} \\ 
  flux\_64 & \textbf{0.938} & -0.102 & -0.041 \\ 
  flux\_256 & \textbf{0.948} & -0.006 & -0.055 \\ 
  flux\_1024 & \textbf{0.956} & 0.078 & -0.032 \\ 
  flux\_batse\_64 & \textbf{0.976} & 0.087 & -0.032 \\ 
  flux\_batse\_256 & \textbf{0.989} & \textbf{0.193} & -0.041 \\ 
  flux\_batse\_1024 & \textbf{0.997} & \textbf{0.247} & 0.007 \\ \hline
\multicolumn{4}{c}{} \\
\hline
 \textbf{Fluence classes} & LD1 & LD2 & LD3 \\ 
  \hline
  t50 & \textbf{0.126} & \textbf{-0.111} & \textbf{-0.427} \\ 
  t90 & \textbf{0.233} & \textbf{-0.112} & \textbf{-0.531} \\ 
  fluence & \textbf{0.950} & \textbf{-0.104} & \textbf{-0.142} \\ 
  fluence\_batse & \textbf{0.960} & -0.040 & \textbf{-0.111} \\ 
  flux\_64 & \textbf{0.815} & \textbf{-0.258} & \textbf{0.422} \\ 
  flux\_256 & \textbf{0.830} & \textbf{-0.223} & \textbf{0.377} \\ 
  flux\_1024 & \textbf{0.865} & \textbf{-0.180} & \textbf{0.301} \\ 
  flux\_batse\_64 & \textbf{0.841} & 0.026 & \textbf{0.378} \\ 
  flux\_batse\_256 & \textbf{0.859} & 0.021 & \textbf{0.339} \\ 
  flux\_batse\_1024 & \textbf{0.890} & 0.029 & \textbf{0.297} \\ \hline
\end{tabular}
\caption{\label{tab:lda_function} The discriminant functions depending on the parameters. We calculated the correlations between the parameters of the GRBs and the discriminant functions. On the top there are the \textit{pflx} data and bottom the \textit{flnc}. It seems that in the \textit{pflx} models the flux is the most important parameter and in the \textit{flnc} models the fluence. It seems that the flux shows more higher correlations which validates Fig.~\ref{fig:lda_good}. We also found better classification in the \textit{pflx}, according to this we can classify better based on the flux than the fluence values. The boldface numbers indicate the significant correlations where the significance was bigger than 5$\sigma$.}
\end{table}

We made the Kruskal-Wallis test between the spectral types with the {\tt kruskal.test()} procedure from R 'stats' package \citep{R}. Table~\ref{tab:kruskal} shows the significance of the variables obtained from the  Kruskal-Wallis rank sum test. It shows that the duration is not a sensitive parameter between the spectral types. We can also see that the Band and smoothly broken power law do not differ based on these parameters. We found there are only three significantly different spectral classes, the Band and the smoothly broken power law groups are much closer to each other.

\begin{table}
\centering
\begin{tabular}{r|rrr}
  \hline
 & T90 & Fluence & Flux 1024 \\ 
 & $[s]$ & $\left[ \frac{erg}{cm^2} \right]$ & $ \left[ \frac{photon}{cm^2\cdot s}\right] $\\
  \hline
  Band & 0.868 & -4.248 & 1.652 \\ 
  Sb. plaw & 0.840 & -4.618 & 1.385 \\ 
  Compt. & 0.780 & -5.189 & 0.870 \\ 
  Plaw & 0.771 & -5.813 & 0.411 \\ \hline
  Sign. level & 0.996 & \textbf{$<2.2\cdot 10^{-16}$} & \textbf{$<2.2\cdot 10^{-16}$}\\
   \hline
\end{tabular}
\caption{\label{tab:kruskal} Means of variables with peak flux, based on classification. We examined the parameters of the GRBs within every \textit{pflx} classes. We found that the durations were the same but the fluence and flux showed significant differences between the classes. To get significances we used the Kruskal-Wallis rank sum test.}
\end{table}

\subsubsection{Reliability of the spectra (cross validation)}

The LDA can also be used as a classification method because every class class a typical value of discriminant function so the method gives the most likely class from the GRBs parameters. We examined connection between the prior (from the database) and posterior (from the LDA) classes. This result is shown in the Contingency Table~\ref{tab:cross_valid}. We found that the relationship between the old and quasi new class is very significant (sign. level < $2.2\cdot 10^{-16}$). It seems that the power law and Comptonized spectral classes can separate well but the smoothly broken power law merge with the Comptonized and Band spectra.

\begin{table}
\centering
\begin{tabular}{r|rrrr}
  \hline
\textbf{Peak flux classes}& \multicolumn{3}{c}{Posterior classes}\\
Prior classes & Band & Sb. plaw & Compt. & Plaw \\ 
  \hline
Band &  44 &  15 &   4 &   0 \\ 
  Sb. plaw &   9 &  16 &   3 &   1 \\ 
  Compt. &  30 & 126 & 527 & 124 \\ 
  Plaw &   1 &   3 &  93 & 846 \\ 
   \hline
\multicolumn{4}{c}{} \\
  \hline
\textbf{Fluence classes}& \multicolumn{3}{c}{Posterior classes}\\
Prior classes & Band & Sb. plaw & Compt. & Plaw \\ 
  \hline
Band & 115 &  30 &  27 &   0 \\ 
  Sb. plaw &  34 &  29 &  36 &   1 \\ 
  Compt. &  92 & 155 & 604 & 278 \\ 
  Plaw &   0 &   3 &  69 & 369 \\ 
   \hline
\end{tabular}
\caption{\label{tab:cross_valid} Contingency table of prior-posterior classes. The rows show what classes have changed the catalog spectra. The columns show the re-classified spectra. We can see that the power law catalog spectra remained power law but many 'new' smoothly broken power law were Comptonized in the catalog and vice versa.}
\end{table}

We saw in Sect.~\ref{ldgb} that the peak flux spectra are better separated than those of the fluence. We tested the reliability of this  conjecture with the bootstrap method.  We generated 1000 bootstrap random samples from the data and we repeated the LDA and calculated the significance of the first discriminant function. We found that the \textit{pflx} type significances were always better and the rate of $\chi^2$ was about 1.5. We examined the reliability of this result and found a very low error probability (p-value~=~$<2.2\cdot 10^{-16}$), so the peak flux models are much better separated.

\subsection{Transition from peak flux to fluence spectral types}

Studying the temporal variation of GRB spectra revealed four different types of behaviors \citep{golenetskii1983,Kumar8_norris,ford1995,veres_meszaros_2012,veres_meszaros_2013,veres_meszaros_2014,yu_evolution,evol_2,evol_3,evol_4_ruffini}:

\begin{itemize}
    \item No measurable change
    \item Gradual softening in course of time
    \item Tracking with the peaks of the light curve
    \item Change but no correlation with the light curve
\end{itemize}

\noindent  In all cases the spectrum of the peak flux is a snapshot of a time varying process. It is evident, therefore, that spectra obtained from the peak flux and the fluence do not necessarily coincide \citep{evol_batse1,evol6,evol_batse2,evol_batse3}.

The difference between the peak flux and fluence spectra can simply be of a statistical nature. In both cases the fitting of a spectral model is based on $\gamma$ photons of a different sample. Namely, the photons responsible for the \textit{pflx} spectra are only a temporal snapshot from the total population resulting fluence spectra, in the duration of the bursts. As we pointed out in Sect.~\ref{ctanal} the transition from peak flux spectral type to the fluence one is not purely accidental \citep{spectral_study,spectral_study2}.

The spectra can be ordered into a series starting at the power law and ending at the Band model. As we show in the next Section this regularity does not necessarily mean physical changes \citep{Kumar18}.

\subsubsection{Role of the signal to noise ratio}

The level of complexity of a fitted model depends on the number of its parameters. The simplest one is the power law model containing only two parameters to fit. It is mostly the only possibility of modeling  faint objects. As the number of the observed $\gamma$ photons is increasing, the possibility of fitting a more complex model is also increasing.

Since the number of $\gamma$ photons of the peak flux is only a fraction of those in the fluence, the transition of a simple model into a more complex one is straightforward. The regularity obtained in Sect.~\ref{ctanal} seems to be explained in this way. There are some concerns, however, about purely accepting this view.

Due to the higher number of photons fitting a model should be more accurate in case of fluence. One may expect a lower statistical uncertainty in this case and a higher one at the peak flux spectra. In the reality, however, the case is just the opposite.

We showed in Sect.~\ref{ldgb} that there is a significant difference among GRBs of different spectral types based on some model-independent physical quantities (e.g. flux, fluence, duration). We demonstrated the strongest contribution to the discriminating variable of Equation~(\ref{ydis}) is given by the flux and the fluence, correspondingly. According to the arguments given above it seems to be quite natural because both quantities relate strongly to the number of the $\gamma$ photons, responsible to the spectra.

\begin{figure}
\centering
\includegraphics[width=4cm]{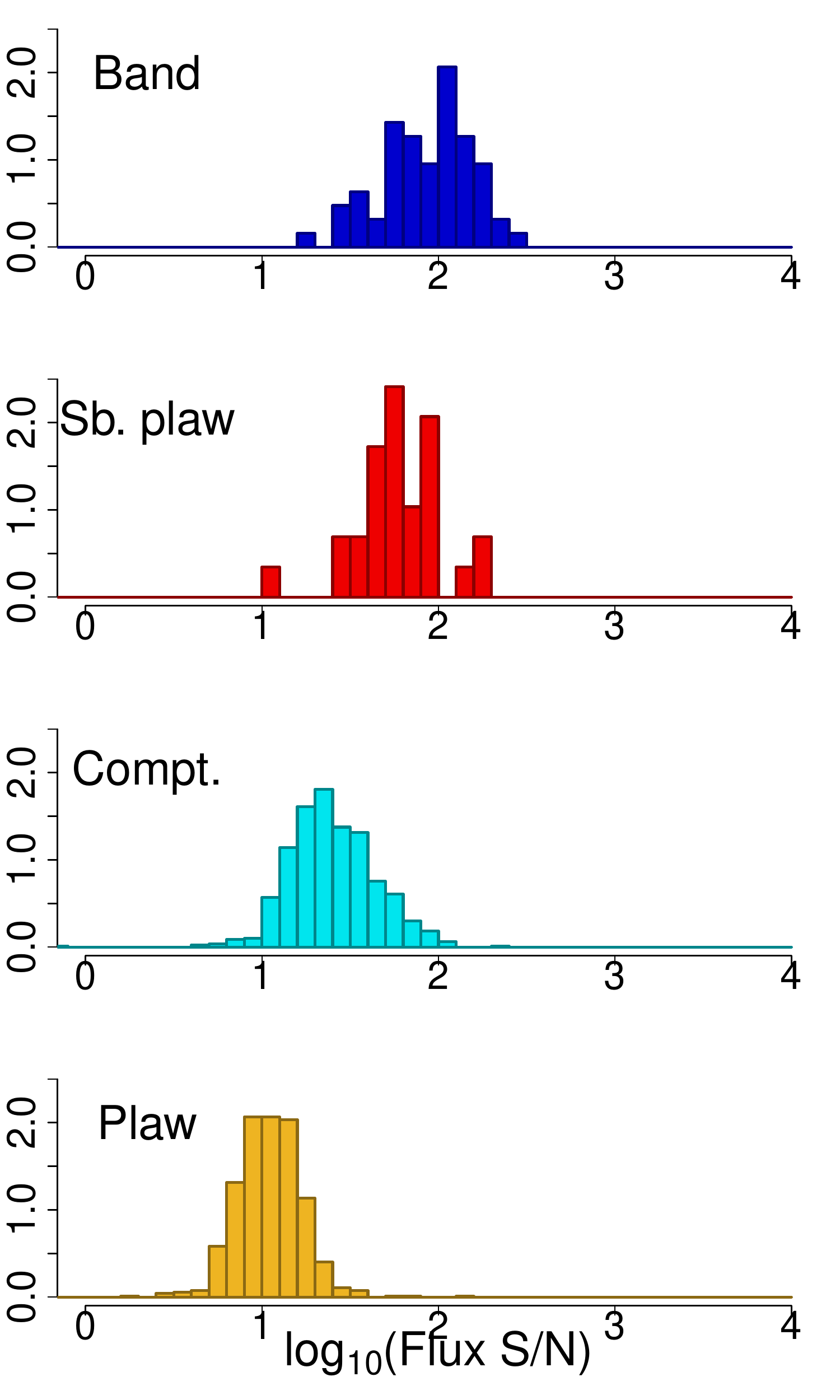}
\includegraphics[width=4cm]{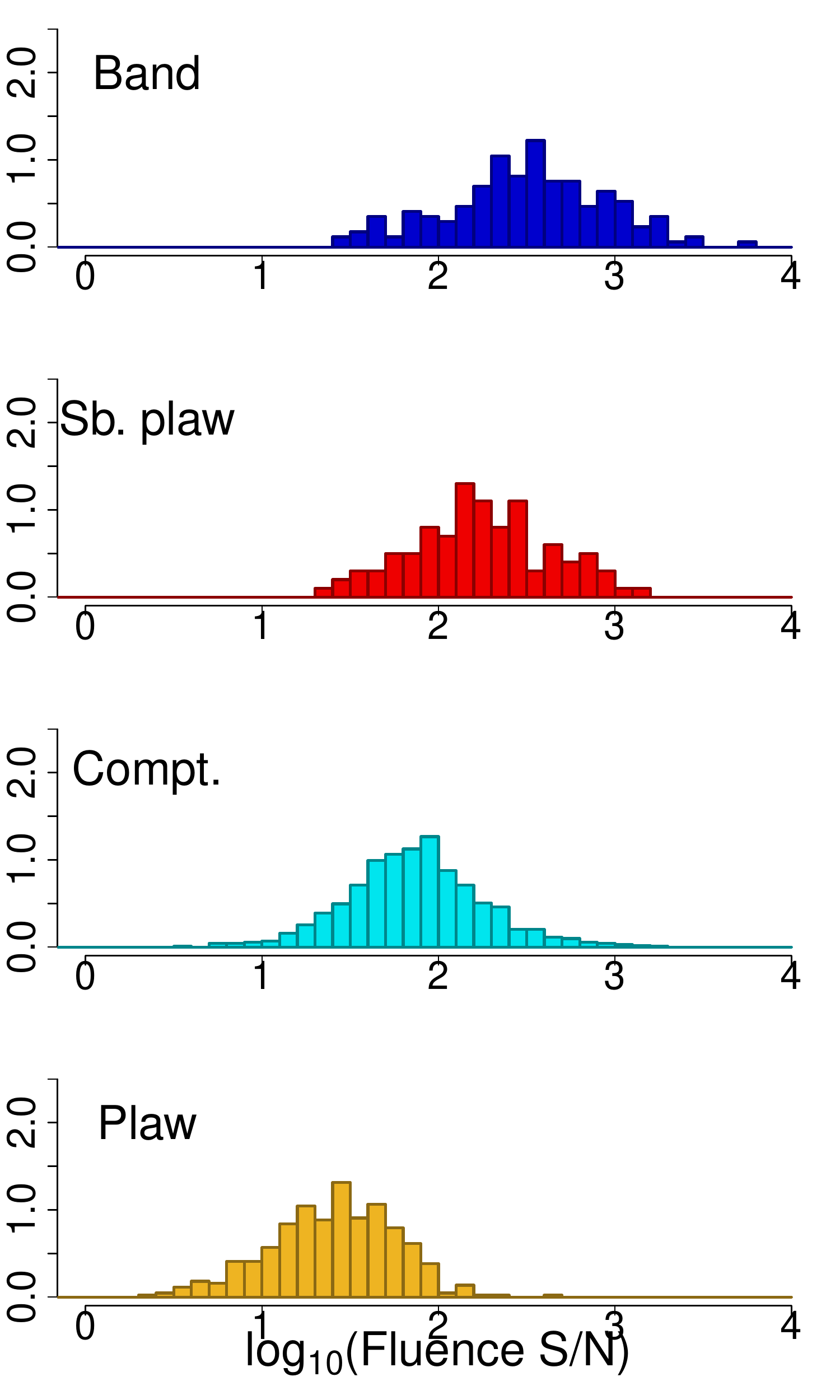}
\caption{\label{fig:sn_full} Distribution of the signal-to-noise ratio within the classes. The fluence S/N distributions are significantly wider than the peak flux within the classes even though there are much more photons in the fluence spectra.}
\end{figure}

The strength of the discriminating power of the $y$ variable can be characterized with the $\lambda$ variable in Equation~(\ref{rbw}). The bigger $\lambda$ is, the stronger its discriminating power. Actually, $\lambda$ is $\sim2$ times higher at the peak flux spectra than those of the fluence. This fact is clearly demonstrated in Fig.~\ref{fig:sn_full}. The $\sigma_w$ variances within the groups are similar in the cases of the peak flux and fluence but the separation between the groups is much more pronounced in the former case. This can be clearly seen in Table~\ref{tab:sn_full} where we demonstrate the flux and fluence averages and the standard deviation within every class. The distance was a little bigger between the average of fluences of classes, but the mean of standard deviation was also much larger (almost double) than the flux.

Despite the higher $\gamma$ photon number of the fluence spectra as compared to the peak flux, its significantly less pronounced discrimination between the classes can be explained by the spectral variation during the outbursts. Namely, the peak flux is a snapshot of a longer time-varying process. Due to the gradual changing  of its  physical properties the fluence spectrum is actually a mixture of different types. Consequently, the fitted spectrum is only a temporal mean of different types.

\subsubsection{Physical differences between the spectral type obtained from the \textit{pflx} and \textit{flnc}}
\label{sec:epeak}

Without temporal variations of the burst spectra the fitted parameters of the peak flux and fluence spectral models should be identical, within the limits of the statistical inference \citep{evol_5}. Due to the temporal spectral changes during the outburst it is not necessarily true.

During an outburst, statistically the smallest change is expected in the fitted parameters when the obtained models are the same in the peak flux and the fluence case. We made comparisons between the model parameters in the case when the best fitting model was identical for both the peak flux and the fluence. We compared two parameters: the peak energy and the low energy spectral index for the cases when it was meaningful (namely, the power law model does not have an energy peak).

Varying SEDs during the outburst result in changes of the fitted parameters \citep{Kumar13,Kumar14,Kumar17}, even if the best fitting model is the same in the peak flux and the fluence case. To get quantitative measures of the eventual changes we compared the mean values of the parameters, performing Student's t tests. The results are summarized in Table~\ref{tab:sn_full}.

\begin{table}
\centering
\begin{tabular}{r|rrr}
  \hline
  & Mean& Mean & Student\\
  & lg(Flux$_{1024}$) & lg(Fluence) & p-value\\
  \hline
  Band & 1.934 & 2.496 & \textbf{$<2.2\cdot 10^{-16}$} \\ 
  Sb. plaw & 1.774 & 2.241 & \textbf{$<2.2\cdot 10^{-16}$} \\ 
  Compt. & 1.398 & 1.866 & \textbf{$<2.2\cdot 10^{-16}$} \\ 
  Plaw & 1.043 & 1.403 & \textbf{$<2.2\cdot 10^{-16}$} \\ 
  \hline
  Max. difference & 0.891 & 1.093 & \\
   \hline
\multicolumn{4}{c}{} \\
\hline
  & St.dev& St.dev & F.test\\
  & lg(Flux$_{1024}$) & lg(Fluence) & p-value\\
  \hline
Band & 0.252 & 0.455 & \textbf{$<2.2\cdot 10^{-16}$} \\ 
  Sb. plaw & 0.243 & 0.386 & 0.006 \\ 
  Compt. & 0.243 & 0.384 & \textbf{$<2.2\cdot 10^{-16}$} \\ 
  Plaw & 0.180 & 0.353 & \textbf{$<2.2\cdot 10^{-16}$} \\ 
  \hline
  Average st.dev & 0.230 & 0.395 & \\
   \hline
\end{tabular}
\caption{\label{tab:sn_full} The mean values of the typical parameters of discriminant functions and between the spectra. We found that the \textit{pflx} was a better classifier where both best models (\textit{pflx} and \textit{flnc}) were similar. The top table shows the means of flux and fluence within the spectral types and classes. The bottom table shows the standard deviations of the spectra. 
It seems that the mean of the fluence is a bit bigger ($\sim$ 10 per cent) than that of the flux. In addition, the \textit{flnc} standard deviations were about two times bigger than \textit{pflx}, so we assume that the flux is a better separator between the classes and this was supported by the significance tests.}
\end{table}

\begin{figure}
\centering
\includegraphics[width=8.25cm]{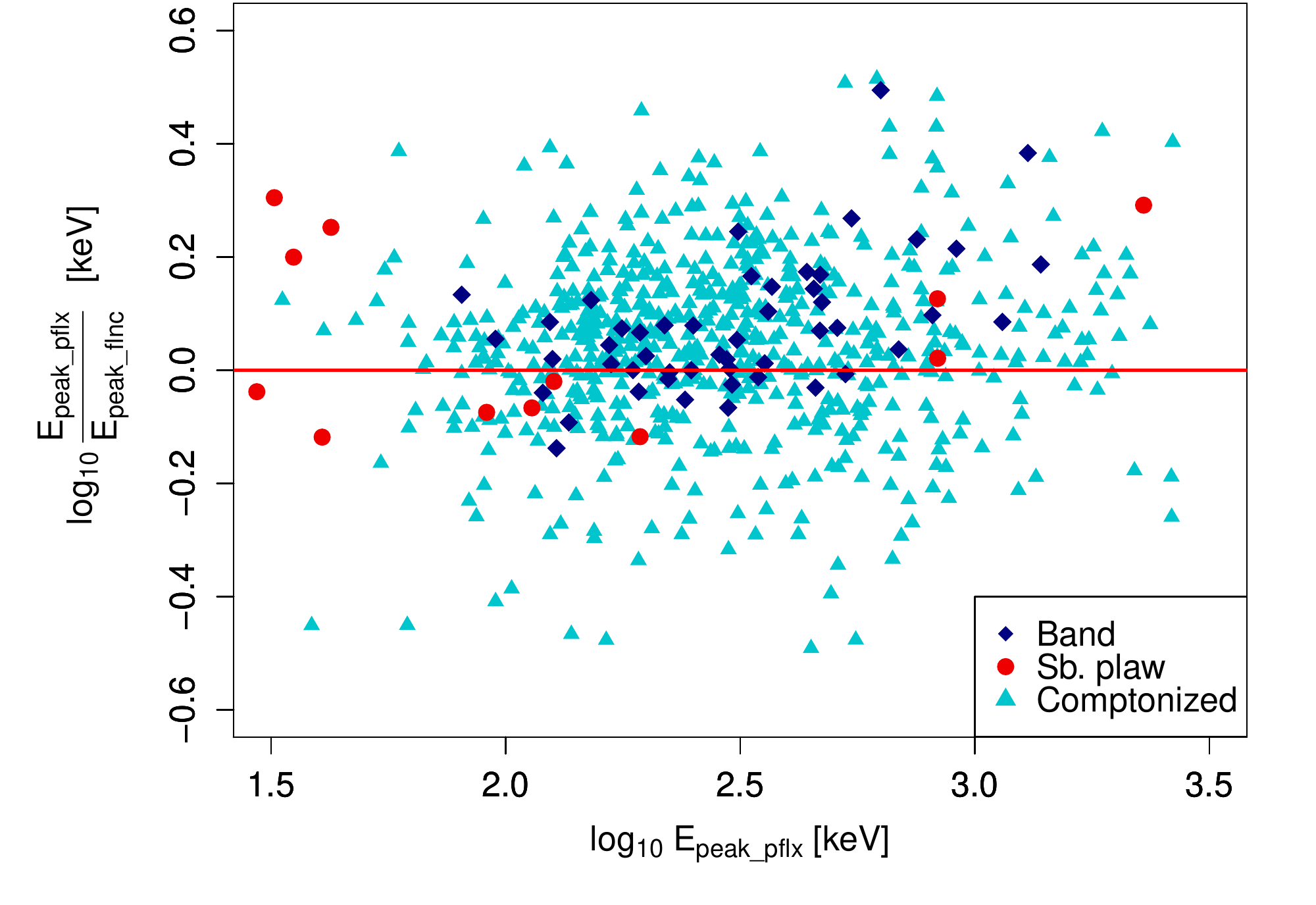}
\caption{\label{fig:evolution} Difference of $\mbox{E}_{\mbox{peak}}$ values for the Band, smoothly broken power law and Comptonized spectra. It can be seen that the $\mbox{E}_{\mbox{peak}}$ of \textit{pflx} type are mostly bigger than the fluence type. It means that we found an evidence for spectral evolution in the $\mbox{E}_{\mbox{peak}}$ values. The blue diamonds show the Band, the red circles the smoothly broken power law, and the turquoise triangles the Comptonized model.}
\end{figure}

Table~\ref{tab:epeak_full} clearly demonstrates that the peak energy of the spectra is softening in the average in course of time \citep{Kumar8_norris, Kumar12}, at the $p=1.4\cdot 10^{-5}$ within the Band, $p=0.1011$  within the sb. plaw, $p=2.0\cdot 10^{-12}$ within the Comptonized spectra level of significance. These results are supported by \citet{ruffini_2004_evol,ruffini_2013_evol,Kumar9,Kumar10}. This result can be also confirmed in Fig.~\ref{fig:evolution} where we plotted the difference of the peak flux and fluence $\mbox{E}_{\mbox{peak}}$ as a function of the peak flux $\mbox{E}_{\mbox{peak}}$. One can infer from the figure that the softening depends on $\mbox{E}_{\mbox{peak}}$ itself \citep{ruffini_2017_evol}. At higher energies it seems to disappear \citep{norris_short2,norris_short1,Kumar19}.

\begin{table}
\centering
\begin{tabular}{r|rrrr}
  \hline
 & Peak flux & Fluence & Sign. & No. of\\ 
 & [keV] & [keV] & level & cases\\
  \hline
  Band & 390 & 306 & $1.4\cdot 10^{-5}$ & 50\\ 
  Sb. plaw & 388 & 280 & 0.101 & 12\\ 
  Comp & 417 & 397 & $2.0\cdot 10^{-12}$ & 620\\ 
   \hline
\end{tabular}
\caption{\label{tab:epeak_full} The means of peak energies of spectra ($\mbox{E}_{\mbox{peak}}$), where the best fitted spectral class were the same in both peak flux and fluence types. It seems that the $\mbox{E}_{\mbox{peak}}$ of \textit{pflx} spectra are bigger than of the \textit{flnc} type, showing that the spectra is softening in time. On the sb.plaw there were only 12 cases so the significance level was not reliable.}
\end{table}

\begin{figure}
\centering
\includegraphics[width=8.25cm]{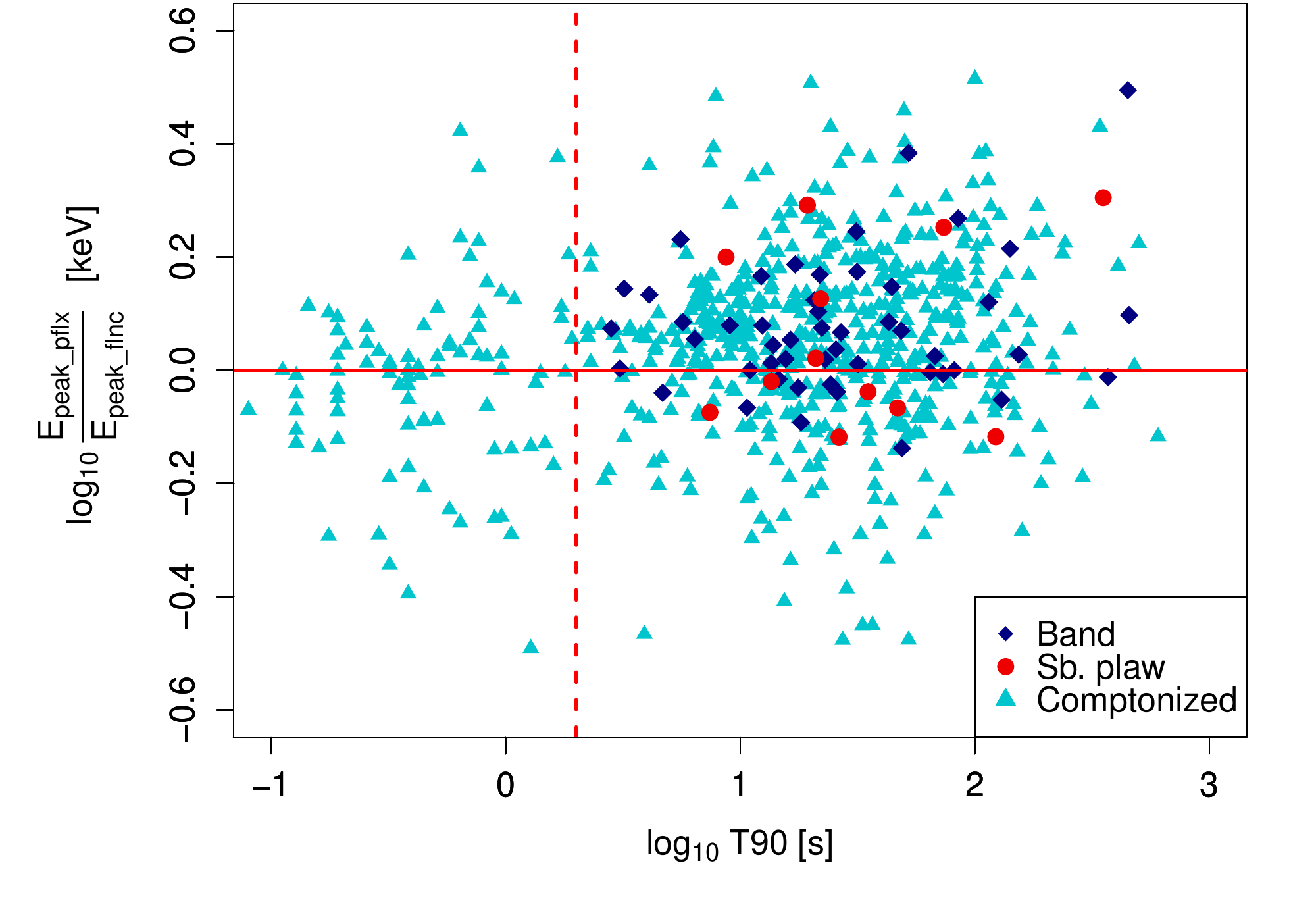}
\caption{\label{fig:evolution_types} Dependence of the $log_{10}\mbox{E}_{\mbox{peak}}$ differences on the $T90$ duration. We compared the peak flux and fluence types $\mbox{E}_{\mbox{peak}}$ from the Comptonized, smoothly broken power law, and Band spectra. We used only those GRBs where the best fitted models were the same in both the peak flux and fluence. We found that the 'long' GRBs usually show spectral evolution and the short GRBs do not. From our data we also could strengthen the fact that the $T90$ limit was around 2 sec. The blue diamonds show the Band, the red circles the smoothly broken power law and the turquoise triangles show the Comptonized model.}
\end{figure}

According to the observations GRBs of short duration \citep{hoi2006,Kumar11_shortgrbs} ($T90 < 2 s$) are harder in the average than long GRBs ($T90 > 2 s$). We displayed in Fig. \ref{fig:evolution_types} the energy difference of $\mbox{E}_{\mbox{peak}}$, defined above, as a function of the $T90$ duration. Seemingly, the average softening of $\mbox{E}_{\mbox{peak}}$ is characteristic only at the long ($T90 > 2 s$) GRBs. It may indicate some real physical differences in the working mode of the short and long duration GRBs. 

Several papers have said that a third group is possible based on the $T90$ distribution \citep{hoi1998,hoi2009,hoi2010,hakkila2003}. Nevertheless, there are some concerns on its existence \citep{tarnopolski2016}. We can see that these GRBs (from the third class) are similar to the long GRBs and these also show the difference of $\mbox{E}_{\mbox{peak}}$.

\begin{figure}
\centering
\includegraphics[width=8cm]{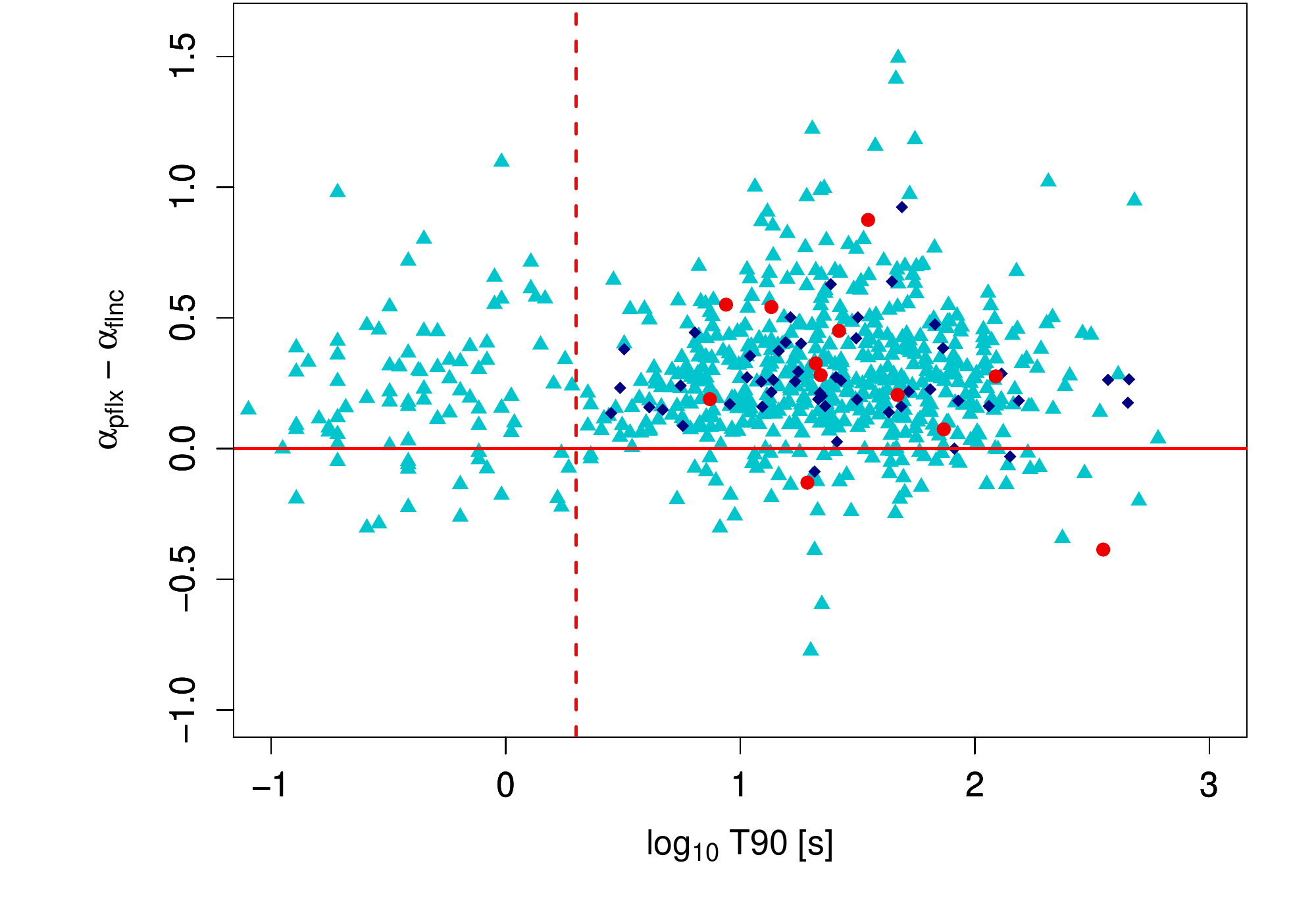}
\includegraphics[width=8cm]{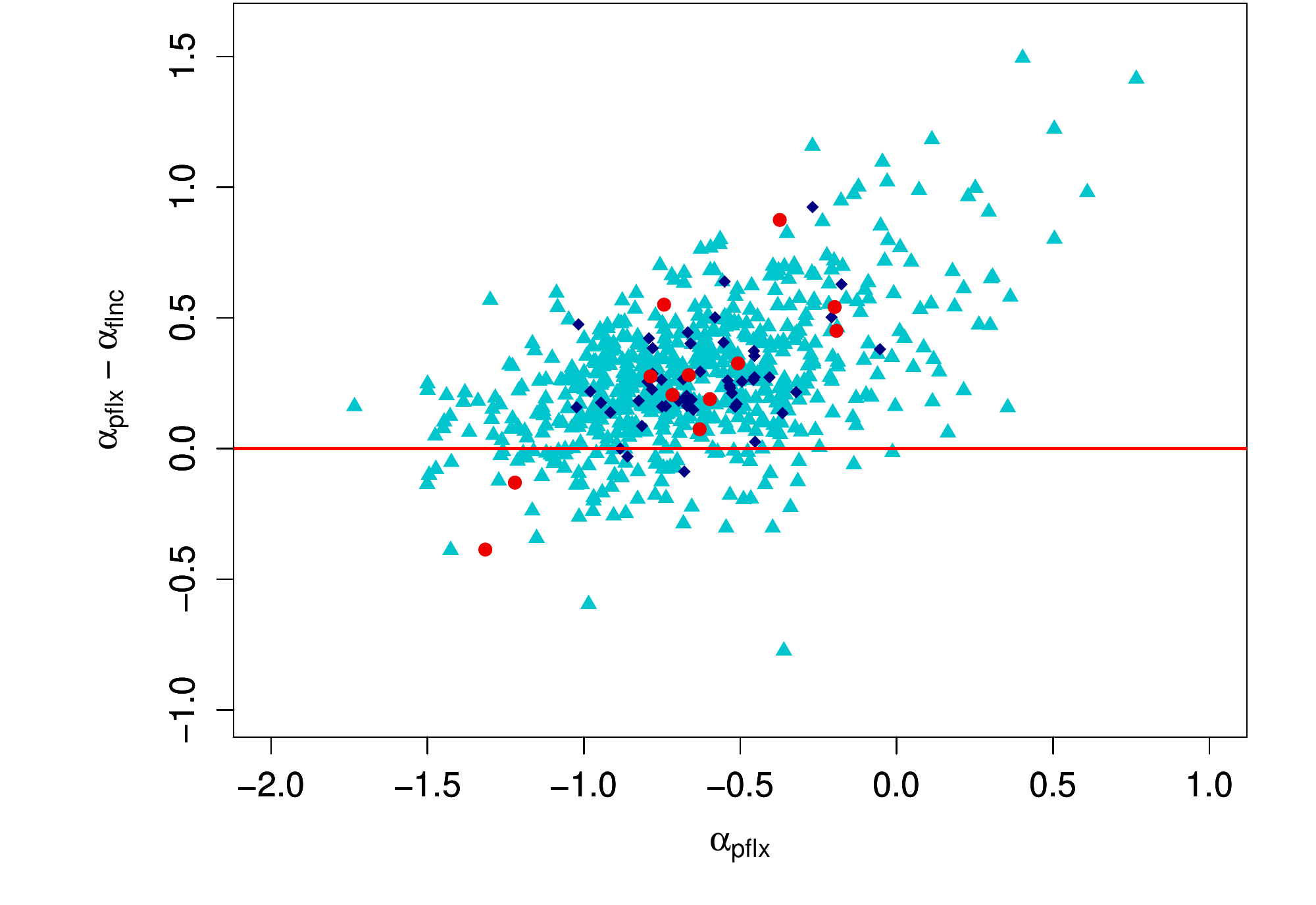}
\caption{\label{fig:evolution_types_lowindex} Evolution of the Low-Energy index. We compared the low energy indices of peak flux and fluence types from the Comptonized, smoothly broken power law, and Band spectra. We used the same GRB subsample as in Fig.~\ref{fig:evolution_types}. Here we can see that both long and short GRBs show significant softening in the spectral index but there is no differences between the short and long GRBs (upper figure). Also, we found a very strong correlation between the evolution of index and the \textit{pflx} low-energy index (lower figure). The notation is the same as in the previous figure.}
\end{figure}

We examined the low energy indices to find evidences for the variation of the spectra \citep{Kumar15, Kumar16}. Our results are seen in Fig.~\ref{fig:evolution_types_lowindex} where we demonstrate how the low energy indices show significant variation. The difference doesn't depend on the duration, so the spectra of both short and long GRBs changed in the same way. The mean difference of low energy indices was 0.21 at the short and 0.28 at the long GRBs which are  significantly different from zero at a very high level. 
In addition, we found a correlation between the low energy index and the difference of index. This correlation was $0.5$ but it was highly significant (p-value~=~$<2.2\cdot 10^{-16}$). It is shown  in the bottom panel of Fig~\ref{fig:evolution_types_lowindex}.

\subsubsection{Comparison of the spectral types detected by the Swift and Fermi}
\label{sec:fermi_swift_disc}

We compared the Fermi and Swift $\gamma$ spectral classes. The Swift defines two spectral groups: (power law and cutoff power law), while there are four classes in Fermi (Band, smoothly broken power law, Comptonized, power law). We found 233 Fermi-Swift pairs where the fitted spectral models were known.  We examined the contingency table of this spectral distribution separately for the Fermi \textit{pflx} and \textit{flnc} types. These contingency tables are shown in Table~\ref{tab:swift_fermi_cont}. Then, similarly to our studies in Sect.~\ref{homtab} we made the full contingency analysis and calculated the Pearson's residuals.

It seems as if there was some relationship between the Swift and Fermi spectra but it is not significant neither with the Pearson's Chi-squared test (p-value~=~0.375) nor with the Fisher's Exact test (p-value~=~0.318). We think that the sample size is too small to get a firm a conclusion from it; there are too few elements in each group (see the row of sb. plaw), therefore we could not evince strong connection between the spectral classes of Swift and Fermi.

\begin{table}
\centering
\begin{tabular}{r|rr}
  \hline
  \multicolumn{3}{c}{\textbf{Peak flux classes}}\\
  Fermi & \multicolumn{2}{c}{Swift}\\
  & Plaw & Cutoff plaw
  \\
  \hline
  Band &  10 &   0 \\ 
  Sb. plaw &   2 &   1 \\ 
  Comp &  87 &  21 \\ 
  Plaw &  94 &  18 \\ 
   \hline
\multicolumn{3}{c}{} \\
\hline
  \multicolumn{3}{c}{\textbf{Fluence classes}}\\
  Fermi & \multicolumn{2}{c}{Swift}\\
  & Plaw & Cutoff plaw
  \\
  \hline
  Band &  19 &   7 \\ 
  Sb. plaw &  12 &   4 \\ 
  Comp & 112 &  23 \\ 
  Plaw &  50 &   6 \\ 
   \hline
\end{tabular}
\caption{\label{tab:swift_fermi_cont} Contingency table of the Swift BAT and Fermi GBM classes. We did not find significant relationship between the Swift gamma spectral types and the Fermi GBM best spectral models.}
\end{table}

\begin{figure}
\centering
\includegraphics[width=8cm]{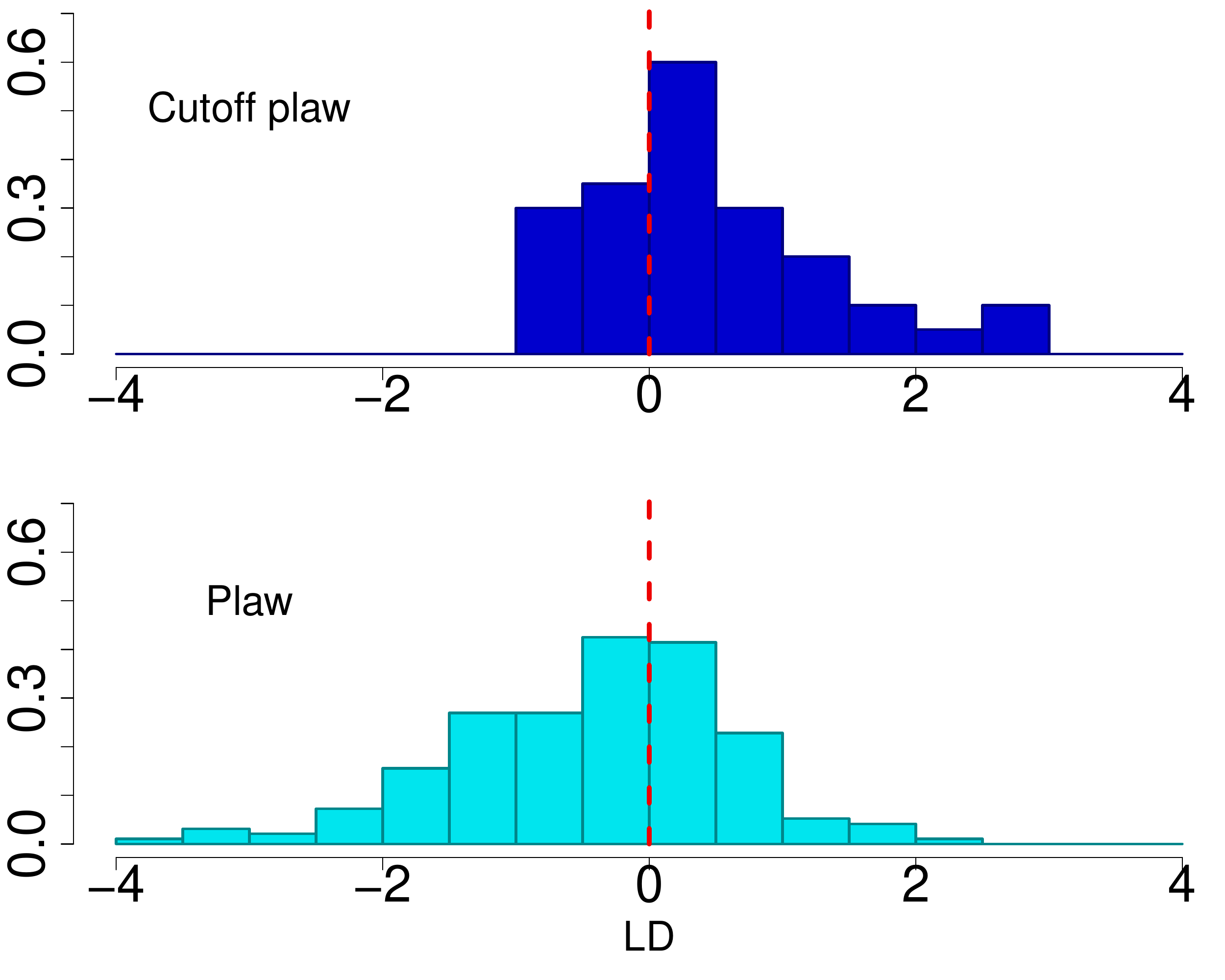}
\caption{\label{fig:swift_lda_hist} The distribution of discriminant function of Swift BAT spectra classes. There seems to be some differences between the distributions but it is only marginally significant (p=0.031).}
\end{figure}

Furthermore, we examined the Swift spectral classes with the LDA method. We used the same Fermi GBM parameters (durations, fluences and fluxes) and we selected only the Fermi best fitted models to the Swift classes. Since here there are only two groups (power law and cutoff power law), we had only one discriminant function. We found here that the significance of the discriminant function was merely $0.031$, which is much weaker than what we experienced above. The distribution of this function is shown in Fig.~\ref{fig:swift_lda_hist}. The Swift $\gamma$ spectra and the Fermi GRB's physical variables.

The correlations between the LD function and the Fermi GBM parameters are in Table~\ref{tab:lda_swift_cor}. One can see that the parameter having the highest correlation is the fluence. Since the Swift spectral classification happens on the full-time spectra like the Fermi \textit{flnc} type, it can explain the correlation with the fluence.

\begin{table}
\centering
\begin{tabular}{r|r}
  \hline
  & LD1 \\ 
  \hline
  t50 & \textbf{0.23} \\ 
  t90 & \textbf{0.34} \\ 
  fluence & \textbf{0.46} \\ 
  fluence\_batse & \textbf{0.50} \\ 
  flux\_64 & \textbf{0.30} \\ 
  flux\_256 & \textbf{0.31} \\ 
  flux\_1024 & \textbf{0.35} \\ 
  flux\_batse\_64 & 0.13 \\ 
  flux\_batse\_256 & 0.19 \\ 
  flux\_batse\_1024 & \textbf{0.25} \\ 
   \hline
\end{tabular}
\caption{\label{tab:lda_swift_cor} Relationship between the discriminant function and the physical parameters of GRBs detected by both satellites. The table shows the result of the LDA and the correlation of the discriminant function with the Fermi model independent parameters. These correlations are mostly significant but smaller than those of the Fermi classes (see Table~\ref{tab:lda_function}). The numbers in boldface indicates the significance levels better than $3\sigma$.}
\end{table}

\subsubsection{Physics of the transition of peak flux to fluence spectra}

According to the widely accepted fireball model \citep{meszaros_fireball,fireball_piran} the huge amount of GRBs' energy (may reach even $10^{54}$ergs) is liberated in a volume  with a diameter of only a couple of ten kilometers, in some cases only within milliseconds.

The byproduct of this suddenly released tremendous energy is a high temperature fireball ($kT \geq MeV$) consisting of $e^\pm$, $\gamma$-rays and baryons but having only a small fraction ($10^{50}-10^{52} erg$) of the total liberated energy \citep{Kumar20}.

The very high optical depth, however, traps the fireball photons, $e^\pm$, nucleons and magnetic fields. The trapped thermal energy is resulted in an accelerating expansion of the fireball, and the thermal is converted into kinetic energy. According to the observations, however, this relativistic expansion occurs along a collimated jet, rather than isotropically \citep{fireball_gehrels}.

In this collimated jet three regions are predicted where the physical conditions become well suited for producing the observed gamma photons. The closest one to the center of the fireball is the photosphere, the boundary of the optically thick central region, at a distance of about $\sim 10^{11}\, cm$ \citep{Kumar}. The photospheric emission has a quasi thermal spectrum.

In the relativistic, baryonic outflow where the Lorentz factor (LF) is time dependent the faster part of the outflow will collide with a slower moving part ahead of it \citep{meszaros_2006}. The collision produces a pair of shock waves and a fraction of jet kinetic energy is converted to thermal energy. The energy is radiated away via synchrotron and the inverse-Compton processes. This internal shock region where these processes are going on locates at a distance of about $\sim 10^{13}\, cm$ from the center of the explosion.

The outflow jet encounters the surrounding interstellar matter at a distance of about $\sim 10^{16}\, cm$ triggering forward/reverse shocks giving rise to a variety of non-thermal spectra.

The synchrotron radiation probably plays a significant role in producing $\gamma$ photons detected by Fermi GBM. A characteristic property of this type of nonthermal radiation that is the low energy part of the intensity distribution is proportional to $E^{1/3}$, independently from the strengths of the magnetic field. Correspondingly, the photon  number distribution is proportional to $E^{\alpha}$, where the power is $\alpha=(1/3-1)=-2/3$ in this case \citep{Rybicki_BOOK}.

Computing the mean value of the $\alpha$ low energy power for the Band, smoothly broken power law and Comptonized peak flux spectra we obtained $\alpha=-0.624\pm0.031$, $-0.662\pm0.100$, $-0.648\pm 0.015$, values, respectively. All of them are equal to  the $\alpha=-0.667$ canonical value, within the limits of the statistical uncertainty. This result indicates that the synchrotron radiation plays a fundamental role in generating $\gamma$ photons observed in the low energy range of the peak flux spectra.

We made a similar computation also for the fluence spectra. The computation resulted in $\alpha=-0.891\pm0.031$, $ -0.933\pm 0.060$, $-0.917\pm0.013$, respectively. Seemingly, all these $\alpha$ values are smaller than the canonical one of the synchrotron radiation. This deviation significantly exceeds the limits of the statistical uncertainty (the probability that the difference is purely accidental is only $p < 2.2\cdot10^{-16}$). Several papers show the same results where the Low-Energy index $\sim -1\pm1$ from Sect.~6.2.1 of \citet{Kumar} or see \citet{thermal1,spectral_study2}. But now it is clear these results show the full time spectra and we found the GRB spectra around the peak point is significantly different.

The canonical value of the low energy spectral index is valid only in the case if the synchrotron cooling is negligible. The cooling makes the low energy part of the spectra softer \citep{cooling}. The difference in he strength of the cooling is an appealing explanation for the difference between the mean low energy spectral index of the peak flux and the fluence spectra. It should be noted, however, that $\alpha = -1$ can also be obtained from the adiabatic non-resonant acceleration in magnetic turbulence in the prompt emission of GRBs \citep{cooling_Xu}.

In case of synchrotron radiation the peak energy of the intensity spectrum depends on the strengths of the magnetic field. Due to the collisions inside the internal shock region the compression of the outflow plasma results in strengthening  the frozen in magnetic field. Assuming that the peak flux is associated with such a compression the strengthening of the magnetic field results in a higher peak energy.

We computed the mean peak energy in case of the Band, smoothed power law and Comptonized spectra for the peak flux and fluence, respectively. The computations resulted in Table~\ref{tab:epeak_full}. As we wrote in the Sect.~\ref{sec:epeak} the long GRBs show that the fluence spectra have systematically lower peak energy and the difference is significant at the $p= 3.39\cdot 10^{-6}$ level.

As one sees in Figure~\ref{fig:evolution_types} the significant difference in the peak energy is valid only for the bursts of $T90>2 \, s$. A trivial explanation would be the difference in the ratio of the time bin of the peak flux to the total duration of the burst. In the reality, however, the time bin is only $64 \,ms$ for the bursts of $T90< \,2s$ and $1 s$ for those of $T90> \,2s$. Instead, the difference may indicate difference between the outflow dynamics of the short and long bursts.

\section{Summary and conclusions}
\label{sucon}

In this paper we analyzed the Fermi GBM Burst Catalog which has two categories of data (\textit{pflx} and \textit{flnc}).
There are four model spectra within the two types: Band, smoothly broken power law, Comptonized and power law. The statistically favored model was selected on both the \textit{pflx} and \textit{flnc} spectra, called '(Pflx/Flnc) Best Fitting Model'. In this paper we analyzed the relationship between the general physical parameters and the best fitting spectral models.

To get the relationship between the best fitting peak flux and fluence spectra we analyzed the statistical properties of their contingency table. 
In the contingency table between the \textit{pflx} and \textit{flnc} types we calculated the row, column and average profiles. Then we determined the difference of the row and column profiles from the average ones. We found that the Comptonized spectra matched best the average profile. Furthermore, we computed the  Pearson's residuals where we can infer from Fig.~\ref{ch2res} that the power law and Band spectra have the strongest deviation.

We did CA using the R $FactoMineR$ library, and Fig.~\ref{CAmap} shows the results. In this figure we can see that the greatest difference was between the power law and Band spectra in both spectra types (peak flux and fluence). In addition, we also got a sequence for the spectra: Band - smoothly broken power law - Comptonized - power law.

To get a relationship between the best fitting model spectra and the general physical parameters we used Fisher's LDA \citep{LDA1,LDA4}. 
Using LDA we studied the relationship between the best fit spectral model and the GRBs' model independent physical parameters. Taking into account all the data where the best model was available we found that at least the first discriminate function was significant on both the peak flux and fluence spectral types. 

We also studied  the 'GOOD sample' according to the criteria published by \citet{fermi1}. We got an improved significance of several magnitudes. The second discriminant functions also became significant here (the p-values~=~$8.5\cdot 10^{-4}$ for the \textit{pflx} and $4.1\cdot 10^{-6}$ for the \textit{flnc}). The resulted separation of classes is shown in Fig.~\ref{fig:lda_good}. Computing the correlations between the discriminant functions and the input variables, the structure matrix, we found that the flux and fluence were important parameters for the discrimination but the durations were not. According to LDA for the \textit{pflx} type the Flux and for the \textit{flnc} type the Fluence is the strongest separator variable.

Although the fluence spectra were fitted with more photons than the peak flux ones - since they were accumulated within the entire burst duration - we obtained the Flux, i.e. the \textit{pflx}, spectra were significantly better separated in the parameter space of the model independent variables. We think this effect can be explained by the variation of the spectra during the outburst. Namely, the interval of the peak flux is only a snapshot of a time varying process responsible for the fluence. 

To study the time variation of the \textit{plfx} and \textit{flnc} spectra quantitatively we compared the Low-energy spectral indices and the $\mbox{E}_{\mbox{peak}}$ values.  Fig.~\ref{fig:evolution_types} shows that there is a difference between the short and long GRB's evolution. Fig.~\ref{fig:evolution} shows a 'hard-to-soft' evolution, as the $\Delta \mbox{E}_{\mbox{peak}}$ (pflx-flnc) are significantly positive. The $\mbox{E}_{\mbox{peak}}$ varied at the long GRBs but not at the short GRBs.

Similar duration effect was not detected in the Low-energy index. It varied significantly at both types of the spectra, but irrespective of the duration. On the other hand we found that there was a significant (p-value~=~$<2.2\cdot 10^{-16}$) correlation ($0.5$) between the differences of the indices and the indices themselves. It means that the harder (flatter) the spectra, the greater are their changes.

We estimated the mean value of the low energy spectral index for the Band, smoothed broken power law and the Comptonized spectra. We obtained a mean value of $\alpha= -0.624\pm0.031, -0.662\pm0.100, -0.648\pm0.015 $ for the \textit{pflx} spectra, respectively. We concluded that the synchrotron radiation is very dominant in case of the peak flux because the obtained mean of low energy indices are with the theoretical value of $\alpha=-0.667$, within the limits of the statistical inference. 

In the case of the \textit{flnc} spectra, however, these values are $\alpha= -0.891\pm0.031 , -0.933\pm0.060, -0.917\pm0.013$, respectively. These values are significantly smaller than the theoretical value of the synchrotron radiation (if cooling can be ignored), the difference shows that the cooling is important in case of the fluence spectra.  

Finally, we compared the Swift and Fermi GRBs, and we found 292 common pairs. We used the LDA to investigate whether there is any relationship between the two Swift gamma spectral classes (power law and cutoff power law) and the Fermi model independent parameters. We found a weak, marginally significant result, a possible connection, but unlike the Fermi classes there is no dominant variable responsible for the differences between the Swift classes.

We tried to find a relation between the GRB spectra and their rest-frame model independent parameters. We got 74 Fermi-Swift GRBs with spectroscopic redshift and we transformed the parameters to the comoving system following \citet{meszaros_cosmo_calc}. We recognized significant differences between the spectra but with a much weaker significance than in the observers frame. The strength of the discriminant power was decreased because the spectra were fitted to the observed photon distribution and this result  was very much influenced by a strong selection effect.

\section*{Acknowledgements}

This work was supported by the Hungarian OTKA NN-111016 grant. Supported through the Hungarian New National Excellence Program of the Ministry of Human Capacities (IR). We are grateful to Rebeka B\H{o}gner and S\'andor Pint\'er for their help in preparing the manuscript. We are thankful to the Anonymous Referee for his helpful suggestions concerning the presentation of this paper.



%\bibliographystyle{mnras}
\bibliographystyle{mnras}
\bibliography{bib}

%%%%%%%%%%%%%%%%%%%%%%%%%%%%%%%%%%%%%%%%%%%%%%%%%%

%%%%%%%%%%%%%%%%% APPENDICES %%%%%%%%%%%%%%%%%%%%%

%\appendix



%%%%%%%%%%%%%%%%%%%%%%%%%%%%%%%%%%%%%%%%%%%%%%%%%%
%%%%%%%%%%%%%%%%%%%%%%%%%%%%%%%%%%%%%%%%%%%%%%%%%%

%%%%%%%%%%%%%%%%%%%%%%%%%%%%%%%%%%%%%%%%%%%%%%%%%%

% Don't change these lines
\bsp	% typesetting comment
\label{lastpage}
\end{document}